\documentclass[]{aa}
\pdfoutput=1
\usepackage{geometry} 
\usepackage{graphicx}
\usepackage{placeins}
\usepackage{afterpage}
	
\usepackage[varg]{txfonts}
\usepackage{upgreek}
\usepackage{cancel}
\bibpunct{(}{)}{;}{a}{}{,} 
\usepackage{soul}

\newcommand{\radu}{rad m$^{-2}$}

\newcommand{\inv}{$^{-1}$}

\usepackage{microtype}
\title{Diffuse polarized emission in the LOFAR Two-meter Sky Survey\thanks{Figures in this version have been degraded to comply with arXiv's file size limits. Full quality figures will appear in the A\&A version of this paper. The data presented in this paper will be available at the CDS via http://cdsweb.u-strasbg.fr/cgi-bin/qcat?J/A+A/ after publication.}}
\titlerunning{LOTSS diffuse}
\author{C.L.~Van Eck\inst{1,2} \and 
M.~Haverkorn\inst{2} \and
M.I.R.~Alves\inst{2} \and
R.~Beck\inst{3} \and
P.~Best\inst{4} \and
E.~Carretti\inst{5} \and
K.T.~Chy\.zy\inst{6} \and
T.~En{\ss}lin\inst{7,8} \and
J.S.~Farnes\inst{2,9} \and
K.~Ferri\`{e}re\inst{10} \and
G.~Heald\inst{11} \and
M.~Iacobelli\inst{12} \and
V.~Jeli\'{c}\inst{13} \and
W.~Reich\inst{3} \and
H.J.A.~R\"{o}ttgering\inst{14} \and
D.H.F.M.~Schnitzeler\inst{3} }
\authorrunning{C. Van Eck. et al.}
\institute{Dunlap Institute for Astronomy and Astrophysics, University of Toronto, 50 St. George Street, Toronto, ON M5S 3H4, Canada; 
\email{cameron.van.eck@dunlap.utoronto.ca} \and
Department of Astrophysics/IMAPP, Radboud University, PO Box 9010, 6500 GL Nijmegen, the Netherlands \and 
Max-Planck-Institut f\"{u}r Radioastronomie, Auf dem H\"{u}gel 69, 53121 Bonn, Germany \and 
SUPA, Institute for Astronomy, Royal Observatory, Blackford Hill, Edinburgh, EH9 3HJ, UK \and 
INAF-Osservatorio Astronomico di Cagliari, Via della Scienza 5, I-09047 Selargius (CA), Italy \and 
Astronomical Observatory, Jagiellonian University, ul. Orla 171, 30-244 Krak\'ow, Poland \and 
Max Planck Institute for Astrophysics, Karl-Schwarzschild-Str. 1, 85748 Garching, Germany \and 
Ludwig-Maximilians-Universit\"{a}t M\"{u}nchen, Geschwister-Scholl-Platz 1, 80539, M\"{u}nchen, Germany \and 
Oxford e-Research Centre (OeRC), Keble Road, Oxford OX1 3QG, UK \and 
IRAP, Universit\'{e} de Toulouse, CNRS, 9 avenue du Colonel Roche, BP 44346, 31028, Toulouse Cedex 4, France \and 
CSIRO Astronomy and Space Science, PO Box 1130, Bentley, WA 6102, Australia \and 
ASTRON, the Netherlands Institute for Radio Astronomy, Postbus 2, 7990 AA Dwingeloo, The Netherlands \and 
Ru{\dj}er Bo\v{s}kovi\'{c} Institute, Bijeni\v{c}ka cesta 54, 10000 Zagreb, Croatia \and  
Leiden Observatory, Leiden University, PO Box 9513, 2300 RA Leiden, The Netherlands 
}
\date{} 

\abstract{Faraday tomography allows us to map diffuse polarized synchrotron emission from our Galaxy and use it to interpret the magnetic field in the interstellar medium (ISM). We have applied Faraday tomography to 60 observations from the LOFAR Two-meter Sky Survey (LOTSS) and produced a Faraday depth cube mosaic covering 568 square degrees at high Galactic latitudes, at 4\farcm3 angular resolution and 1 \radu\ Faraday depth resolution, with a typical noise level of 50--100 $\muup$Jy per point spread function (PSF) per rotation measure spread function (RMSF) (40--80 mK RMSF\inv). While parts of the images are strongly affected by instrumental polarization, we observe diffuse polarized emission throughout most of the field, with typical brightness between 1 and 6 K RMSF\inv, and Faraday depths between $-7$ and +25 \radu. 

We observed many new polarization features, some up to 15\degr\ in length. These include two regions with very uniformly structured, linear gradients in the Faraday depth; we measured the steepness of these gradients as 2.6 and 13 \radu\ deg\inv. We also observed a relationship between one of the gradients and an \ion{H}{i} filament in the local ISM. Other ISM tracers were also checked for correlations with our polarization data and none were found, but very little signal was seen in most tracers in this region. We conclude that the LOTSS data are very well suited for Faraday tomography, and that a full-scale survey with all the LOTSS data has the potential to reveal many new Galactic polarization features and map out diffuse Faraday depth structure across the entire northern hemisphere.}


\begin{document}
\maketitle

\section{Introduction}\label{sec:intro}
Magnetic fields are present throughout interstellar space and play an important role in many aspects of the interstellar medium (ISM), such as cloud collapse during star formation \citep{VanLoo12},  energy transports and cascades in magnetohydrodynamic turbulence \citep{Beresnyak15}, and pressure balance between different gas phases \citep{Boulares90}.

Interstellar magnetic fields can be measured using radio polarization through two processes: synchrotron emission and Faraday rotation. Synchrotron emission is produced throughout interstellar space by cosmic-ray electrons as they are accelerated by interstellar magnetic fields, resulting in polarized radio emission. Faraday rotation occurs when polarized emission passes through magnetized plasma (which fills most of the volume of the ISM), which causes a frequency-dependent rotation of the polarization angle. The change in the polarization angle ($\Delta \theta$) is given by 
\begin{equation}
\Delta \theta = \lambda^2\, \phi(d) = \lambda^2 \, \left[ 0.812\; {\rm rad \, m^{-2}} \int_{\mathrm{source}}^{\mathrm{observer}} \left( \frac{n_\mathrm{e}}{{\rm cm^{-3}}}\right) \left( \frac{\vec{B}}{{\rm \upmu G}} \right) \cdot \left(\frac{\vec{dl}}{{\rm pc}} \right) \right],
\end{equation}
where $\phi(d)$ is the Faraday depth which depends on the free electron density ($n_e$) and magnetic field ($\vec{B}$) along the line of sight integrated from the emission source at distance $d$ to the observer.

Polarized synchrotron emission is produced throughout the Galaxy, and then undergoes Faraday rotation as it propagates through the ISM; therefore along any line of sight we see the superposition of the emission at all distances and with correspondingly different Faraday depths. The rotation measure (RM) synthesis technique \citep{Brentjens05} can be used to transform the observed wavelength-dependent polarization into the distribution of polarized emission as a function of Faraday depth. When applied to 3D image-frequency data this is called Faraday tomography, which produces Faraday depth cubes that map out the diffuse polarized emission as a function of position on the sky and Faraday depth. Such observations can be used to constrain magnetic fields in the ISM and study their properties \citep[e.g., ][]{Schnitzeler2007, Jelic15, Lenc16, VanEck17}.

The resolution in Faraday depth of Faraday tomography depends on the range of wavelength-squared sampled by the observations, so low-frequency radio telescopes (which can produce very large wavelength-squared coverage) are capable of achieving much finer Faraday depth resolution than higher-frequency instruments. The newest generation of very low-frequency radio telescopes such as the Low Frequency Array \citep[LOFAR,][]{vanHaarlem2013} and the Murchison Widefield Array  \citep[MWA,][]{Tingay2013}, which operate in the 100--200 MHz range, are capable of reaching Faraday depth resolutions around 1 \radu, approximately two orders of magnitude smaller than the resolution that can be achieved at 1.4 GHz or higher frequencies.

In recent years, there have been many studies using Faraday tomography to study diffuse polarization, each with different observations balancing tradeoffs between field-of-view, angular resolution, and Faraday depth resolution: \citet{Jelic14}, \citet{Jelic15}, and \citet{VanEck17} used LOFAR observations, \citet{Bernardi2013} and \citet{Lenc16} used MWA observations, \citet{Hill17} used data from the Global Magneto-Ionic Medium Survey \citep[GMIMS, ][]{Wolleben2008}, while \citet{MarcoFan} used observations from the Westerbork Synthesis Radio Telescope (WSRT).
The observations with high angular resolution tend to have smaller fields of view, and vice versa, making it difficult to observe the full size of large (several degree) polarization features (such as the filament in \citet{Jelic15} that extends outside the field of view) while still minimizing the effects of beam depolarization (which reduces the signal present in low-resolution observations).


In this paper we report on the polarization processing of 60 LOFAR observations from the LOFAR Two-meter Sky Survey \citep[LOTSS,][]{Shimwell2017} and present a Faraday depth mosaic covering an area of 568 square degrees. This is the first work to combine multiple LOFAR observations into a single Faraday depth cube. This allows us to take advantage of the high angular and Faraday depth resolution of LOFAR while still looking at large-scale features in the diffuse polarization. In Sect. \ref{sec:data} we present our data reduction and the production of Faraday depth cubes for each observation. In Sect. \ref{sec:mosaic} we present a Faraday depth cube mosaic produced from these observations, and highlight the diffuse polarization features seen. In Sect. \ref{sec:tracers} we compare our mosaic against other tracers for components of the ISM, and in Sect. \ref{sec:interpretation} we discuss possible origins for the polarized features we observe. In Sect. \ref{sec:gradients} we consider the depolarization effects of Faraday rotation gradients and the implications for our observations and others. In Sect. \ref{sec:conclusions} we summarize our results.

\section{Data processing}\label{sec:data}
\begin{figure*}[htbp]
   \centering
   \includegraphics[width=\linewidth]{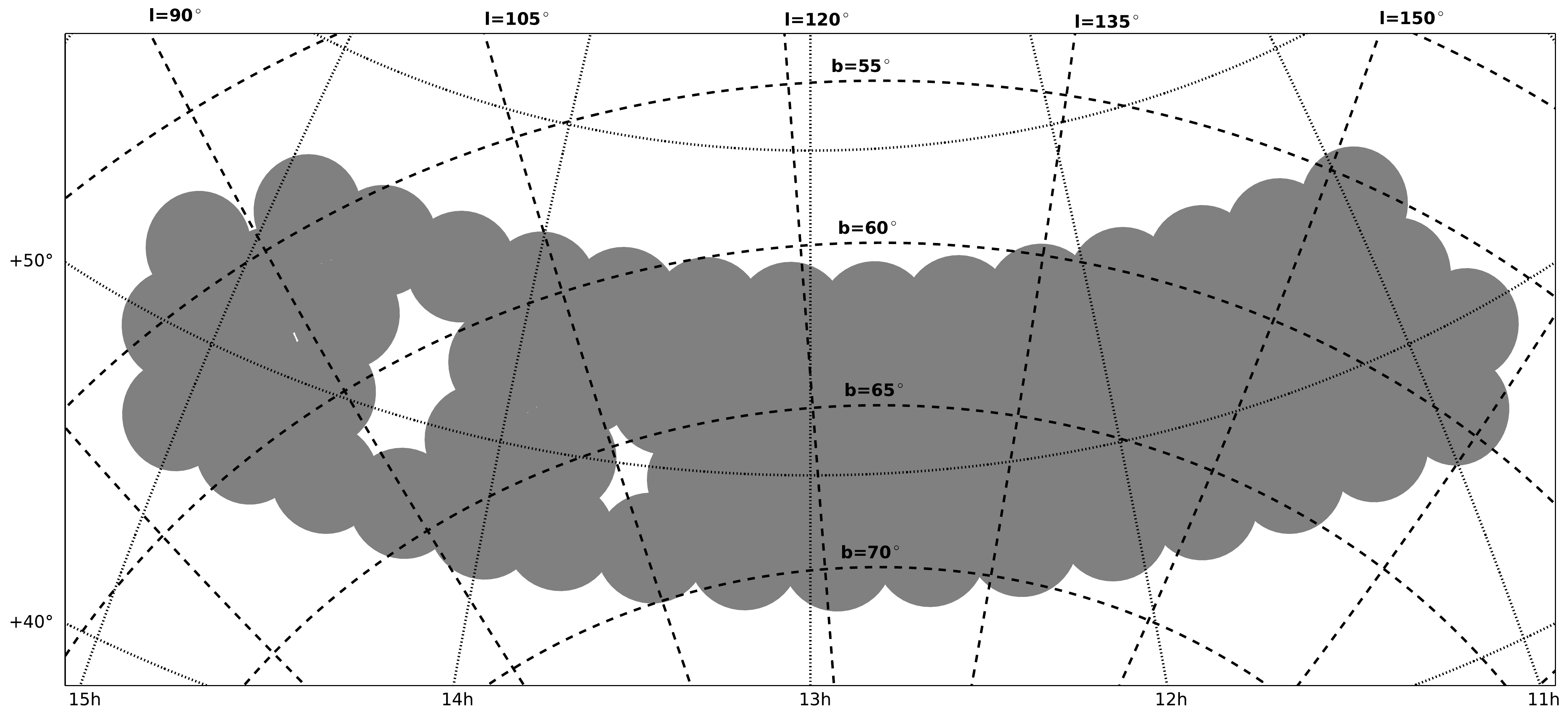} 
   \caption{Locations of the LOTSS HETDEX observations, in both Equatorial (dotted lines) and Galactic (dashed lines) coordinates. Each grey circle covers the area inside the primary beam FWHM (3.4\degr\ diameter) for one observation, at the highest frequency (168 MHz). This image, and all that follow, are in an orthographic projection. The gaps are caused by the three fields that were removed for being very strongly affected by instrumental leakage.}   \label{fig:pointing_map}
\end{figure*}

We analyzed 63 calibrated visibility datasets generated with observations from the LOFAR Two-meter Sky Survey; full details of the observational parameters and calibration methods can be found in \citet{Shimwell2017}. The observations are from the LOTSS test region, right ascension from 10$^\mathrm{h}$30$^\mathrm{m}$ to 15$^\mathrm{h}$30$^\mathrm{m}$ and declination from 45\degr\ to 57\degr, which covers the HETDEX Spring field \citep{Hill08}, a region near the Galactic north pole. Each observation has a nominal duration of 8 hours and covers the frequency range 120--168 MHz with 488 channels. We received the observations after direction-independent amplitude and phase calibration with the LOTSS pipeline. Figure \ref{fig:pointing_map} shows the coverage of these observations in equatorial and Galactic coordinates. The polarization calibration and imaging closely follows that of \citet{VanEck17} and is described below.

Polarization calibration in the form of a correction for ionospheric Faraday rotation was performed prior to imaging. This was done using the RMextract package\footnote{https://github.com/maaijke/RMextract/} written by Maaijke Mevius, combined with maps of the ionospheric total electron content from the Center for Orbit Determination in Europe (CODE)\footnote{http://aiuws.unibe.ch/ionosphere/}. This package produced predictions for the ionospheric Faraday rotation that were used by the Black Board Selfcal software \citep[BBS,][]{Pandey2009} to derotate the polarization of the visibilities and thereby remove the ionospheric Faraday rotation. The estimated error in Faraday depth correction is approximately 0.1 \radu\ \citep{VanEck2018a}. No correction for polarization leakage was performed, as it was deemed too computationally expensive using the currently available methods. The effects of polarization leakage on the results are discussed in Sect. \ref{sec:mosaic_leakage}.

Polarization imaging was performed independently for each channel using AWImager \citep{Tasse13}. A baseline length upper limit of 800$\lambda$ was applied, as this made the synthesized beam (hereafter, point spread function or PSF) more uniform across the band, giving a FWHM of 4\farcm3. The short baselines between the two halves of each high-band antenna (HBA) station \citep[i.e., substations CS(X)HBA0 and CS(X)HBA1, iterating (X) over all HBA stations; see][for a description of the substation layout and naming]{vanHaarlem2013} were removed, as these were known to often suffer from significant mutual interference. A robust weighting of 1.0 was used for imaging, and no CLEANing was performed, as the signal to noise ratio in individual channels was expected to be too low for CLEAN to be effective. By default AWimager produces images both with and without correction for the LOFAR primary beam; the images without correction were used for the quality control step described next, while the images with correction were used for producing Faraday depth cubes.

We found that a small fraction of the images were strongly affected by interference or calibration problems, which usually manifested as very strong patterns throughout the image. To identify these channels in an automated way, we calculated the standard deviation over all pixels in each of the images without primary beam correction (separately for Stokes $Q$ and $U$). The channels affected by these problems stood out as having abnormally large standard deviations, but variations in the background noise across the band and between different observations made it difficult to assign a single threshold value for classifying images as bad. We chose to calculate, for each channel, the median of the standard deviations of the images within 50 channels above or below in frequency; if the standard deviation (for either $Q$ or $U$ separately) in a channel was more than 1.5 times the median standard deviation of these neighbouring channels then that channel was flagged as bad. These bad channels were removed before RM synthesis. During this process, three observations were found to have much higher standard deviations across the full bandwidth than the others, which was due to the presence of extremely bright radio sources; these observations were removed and not included in the following steps. After this step, the typical noise level in a single channel was 2--5 mJy PSF\inv.

RM synthesis was performed using pyrmsynth\footnote{https://github.com/mrbell/pyrmsynth}. Channel weights were calculated using the inverse square of the standard deviation of each channel that was calculated in the previous step. For most observations, the weights were fairly constant across the band, so this weighting choice was not expected to produce any significant changes from uniform weighting. Weights were applied independently per channel; no adjustments for sampling density in $\lambda^2$ were applied. From the frequency coverage and weighting used, the resulting rotation measure spread function (RMSF) had a typical full-width at half-max (FWHM) of 1.2 \radu. From equation 61 of \citet{Brentjens05}, the theoretical FWHM (for uniform weighting) is 1.15 \radu, so these values are consistent and show that the natural weighting has not significantly degraded the Faraday depth resolution. From equations 62 and 63 of \citet{Brentjens05}, our observations are not sensitive to extended structures in Faraday depth wider than about 1.0 \radu, and we have limited sensitivity to Faraday depths larger in absolute value than 170 \radu. Faraday depth measurements beyond this range may also be unreliable \citep{Schnitzeler2015a}. As is typical for LOFAR observations, we are unable to resolve any features with Faraday thickness greater than the resolution, therefore we are only able to pick up unresolved features \citep[or sharp edges of Faraday thick features, as described in][]{VanEck17}. No spectral index correction was included in the RM synthesis step; this may produce minor errors in the polarized intensity and small secondary peaks in the RM spectrum, but these were not expected to affect our analysis. Each RM cube covered Faraday depths from -100 to + 100 \radu \ in steps of 0.25 \radu. The RM-CLEAN algorithm \citep{Heald09} was applied to each Faraday depth cube, to a depth of 2 mJy PSF\inv \ RMSF\inv.

In order to combine the Faraday depth cubes from the different observations, we estimated the position dependent noise in the cubes. We did this on a per-pixel basis by taking the distribution of polarized intensity values taken from the regions of the spectrum that were expected to be dominated by noise ($|\phi| > 20$ \radu) and fitting a Rayleigh distribution (which is the expected distribution for polarized intensity in the absence of signal, \citealt{Macquart2012,Hales2012}). The resulting Rayleigh $\sigma$ parameter (which is equivalent to the Gaussian $\sigma$ of the underlying Stokes $Q$ and $U$ distributions) was taken to be the noise in that pixel.

To merge the separate Faraday depth cubes we averaged the overlapping regions using inverse-variance weighting to produce a final mosaic covering the entire HETDEX region. Alternative weighting schemes, where each pixel used only the observation with the lowest noise or the observation with the nearest pointing center, were also tried but were found to produce significant artifacts in the overlap regions of adjacent pointings. However, these were useful for assessing the quality of the data and the effectiveness of the mosaicing; we observed that many features in the cubes, such as depolarization canals, could be traced continuously across the boundaries between different observations.

\section{The HETDEX mosaic}\label{sec:mosaic}
\begin{figure*}[!hp]
   \centering
   \includegraphics[width=0.9\linewidth,height=0.8\textheight,keepaspectratio]{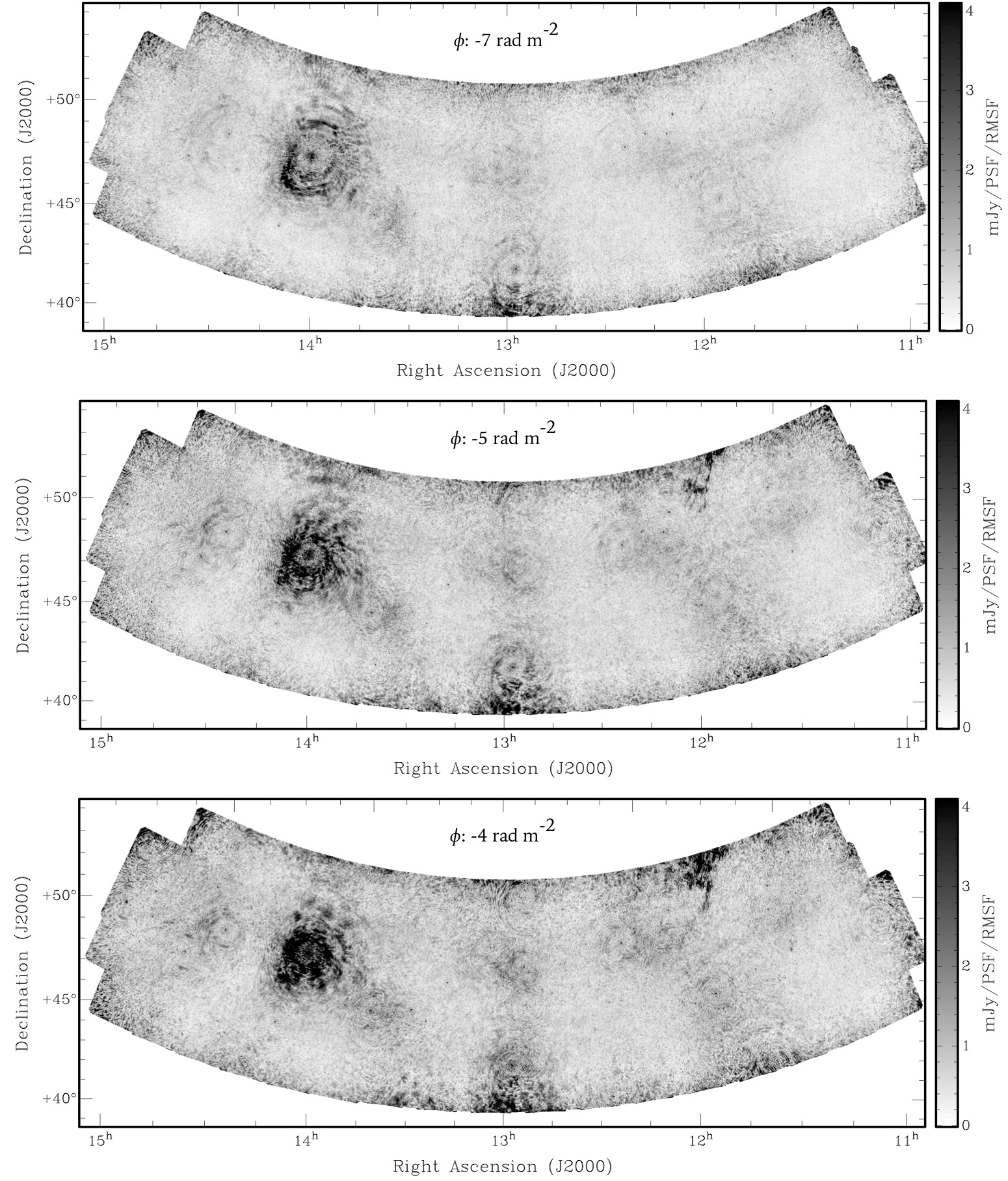} 
   \caption{Selected slices from the Faraday depth cube mosaic, showing the polarized intensity between $-7$ and $-4$ \radu. The top panel shows a typical quiescent slice, with strong artifacts around 3C295. The lower two panels show diffuse polarized emission appearing in the top right.}
   \label{fig:slices1}
\end{figure*}

\begin{figure*}[!p]
   \centering
   \includegraphics[width=\linewidth,height=0.9\textheight,keepaspectratio]{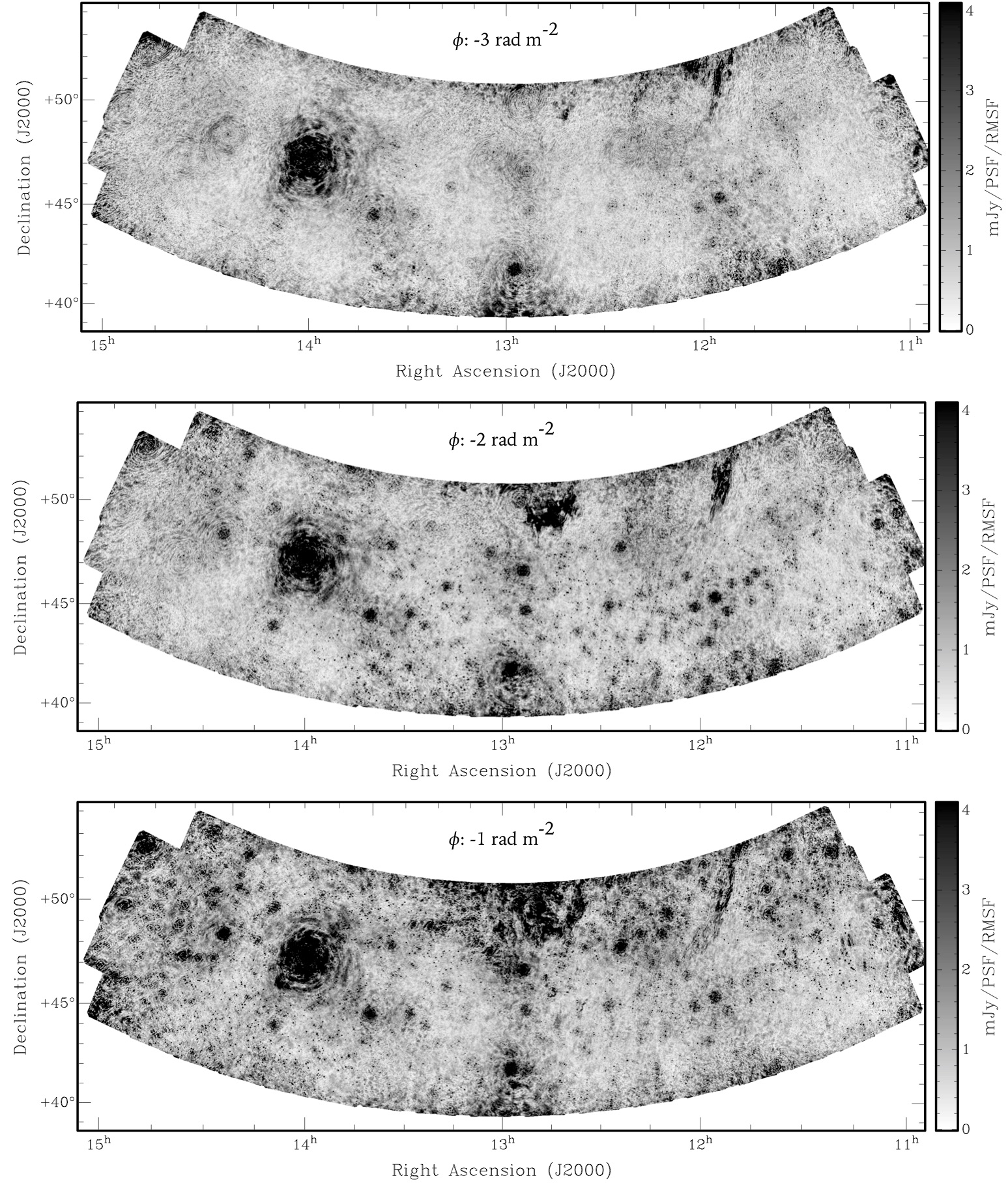} 
   \caption{Selected slices from the Faraday depth cube mosaic, showing the polarized intensity between $-3$ and $-1$ \radu. The feature in the top right travels south-westward, and another feature appears in the top center and spreads outwards. The instrumental leakage from bright Stokes $I$ sources also appear strongly at $-2$ and $-1$ \radu.}
   \label{fig:slices2}
\end{figure*}

\begin{figure*}[!p]
   \centering
   \includegraphics[width=\linewidth,height=0.9\textheight,keepaspectratio]{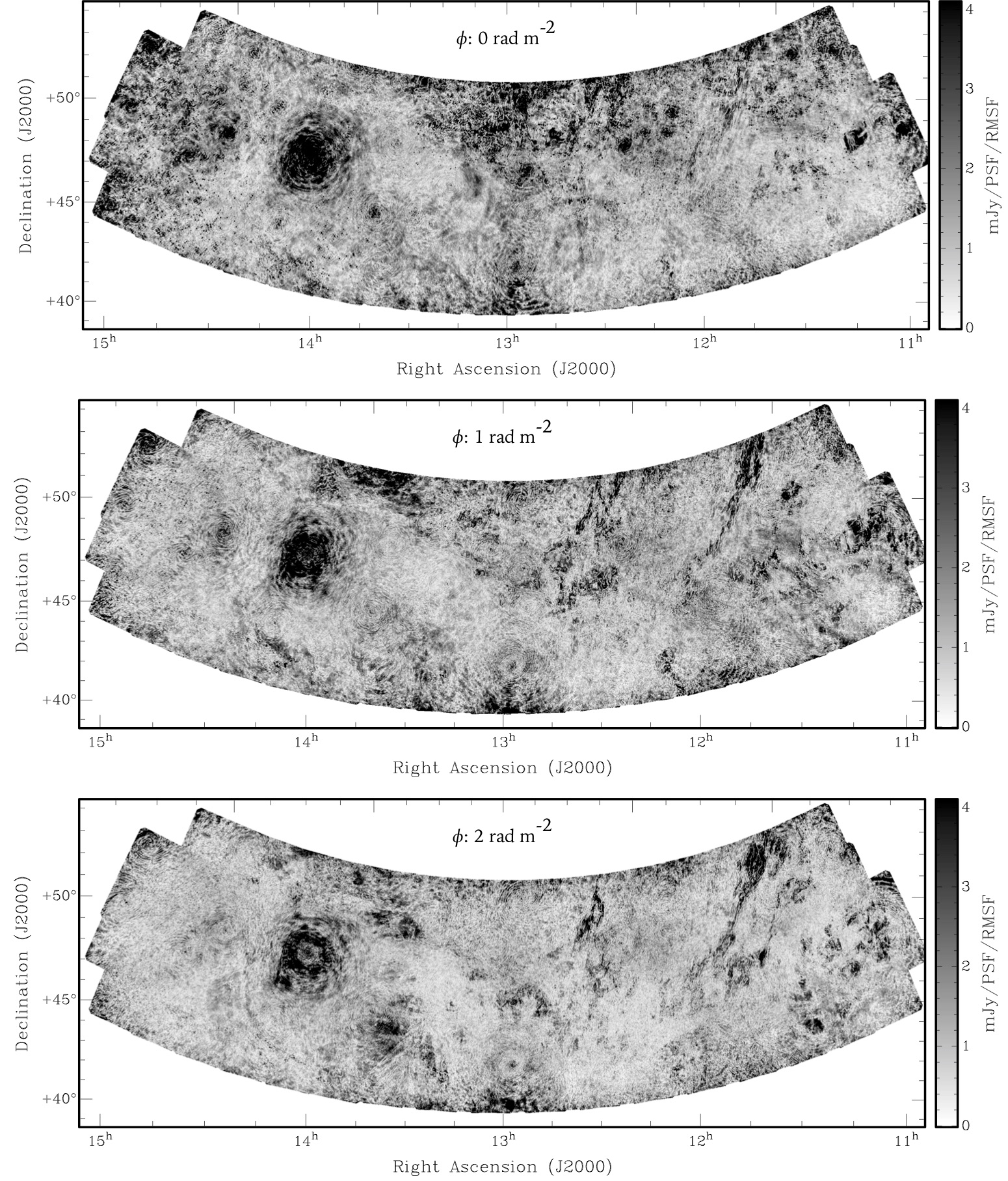} 
   \caption{Selected slices from the Faraday depth cube mosaic, showing the polarized intensity between 0 and 2 \radu. The top-right feature continues to travel south-westward, becoming very filamentary-looking, while the top-center feature continues to expand outwards. Two more diffuse features appear in the lower right corner and lower left-of-center.}
   \label{fig:slices3}
\end{figure*}

\begin{figure*}[!p]
   \centering
   \includegraphics[width=\linewidth]{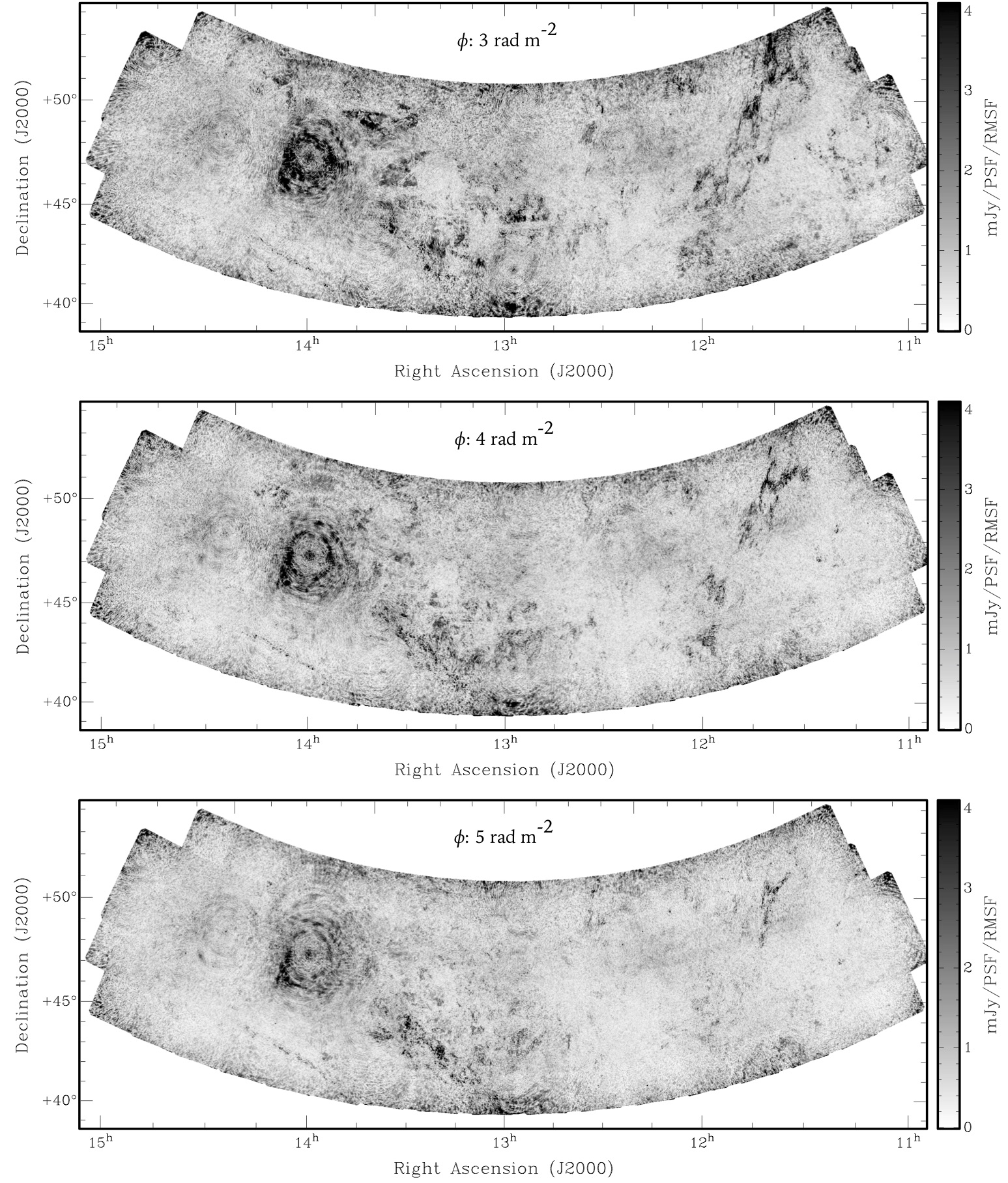} 
   \caption{Selected slices from the Faraday depth cube mosaic, showing the polarized intensity between 3 and 5 \radu. The top right feature now appears as a very long filamentary structure spanning the height of the image, before fading away. The lower right feature also fades away at higher Faraday depths. The top center and lower left-of-center features now appear as a series of patches running south-westward just left of center. A faint, very straight filamentary feature appears in the bottom mid-left, slowly traveling upwards.}
   \label{fig:slices4}
\end{figure*}

\begin{figure*}[!p]
   \centering
   \includegraphics[width=\linewidth]{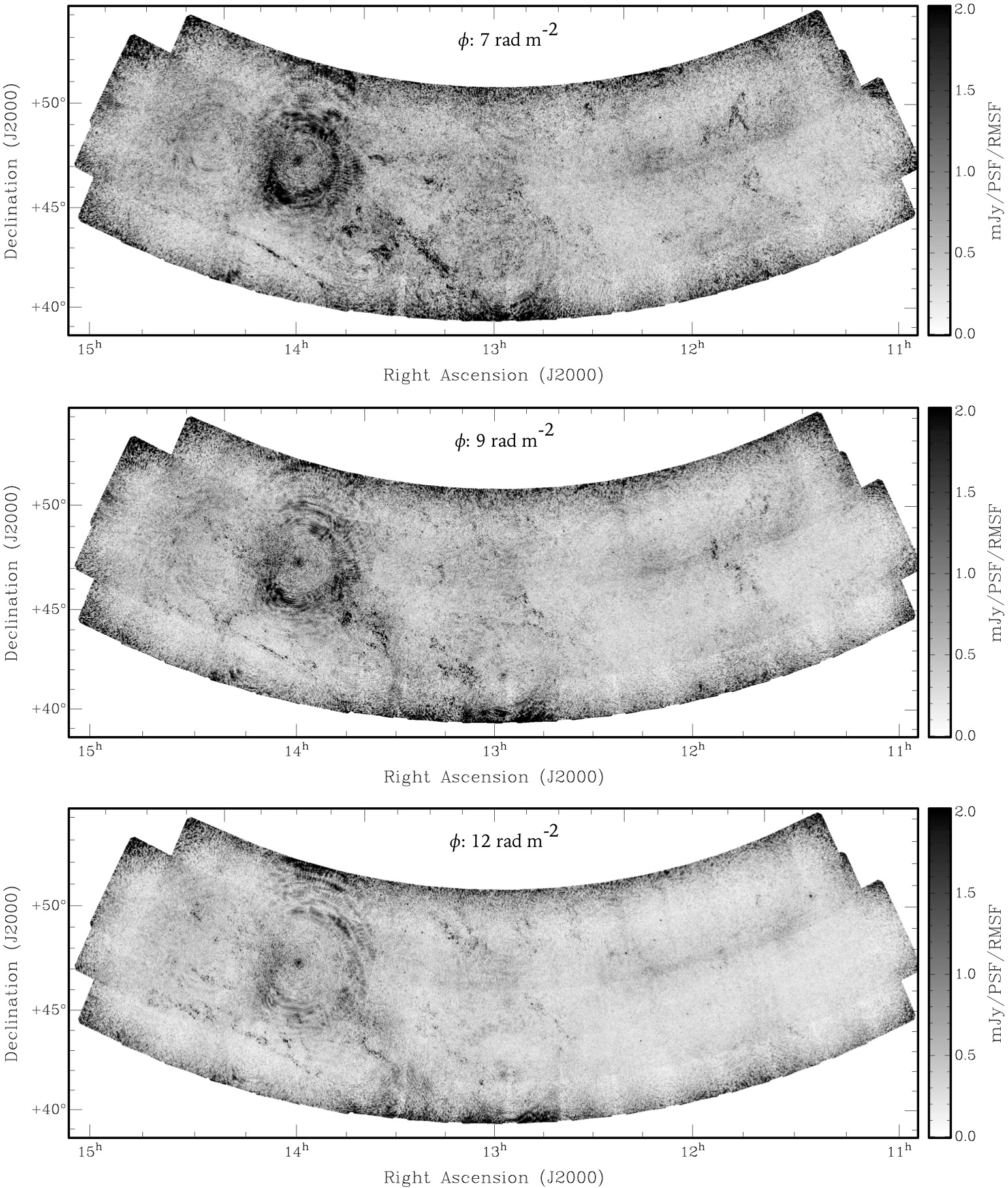} 
   \caption{Selected slices from the Faraday depth cube mosaic, showing the polarized intensity between 7 and 12 \radu. The intensity scale has been adjusted as the emission at these depths is much fainter. The top right and left-of-center features continue to fade out, while the faint very straight filament continues to travel upwards with new, thin features appearing above it.}
   \label{fig:slices5}
\end{figure*}

\begin{figure*}[!p]
   \centering
   \includegraphics[width=\linewidth]{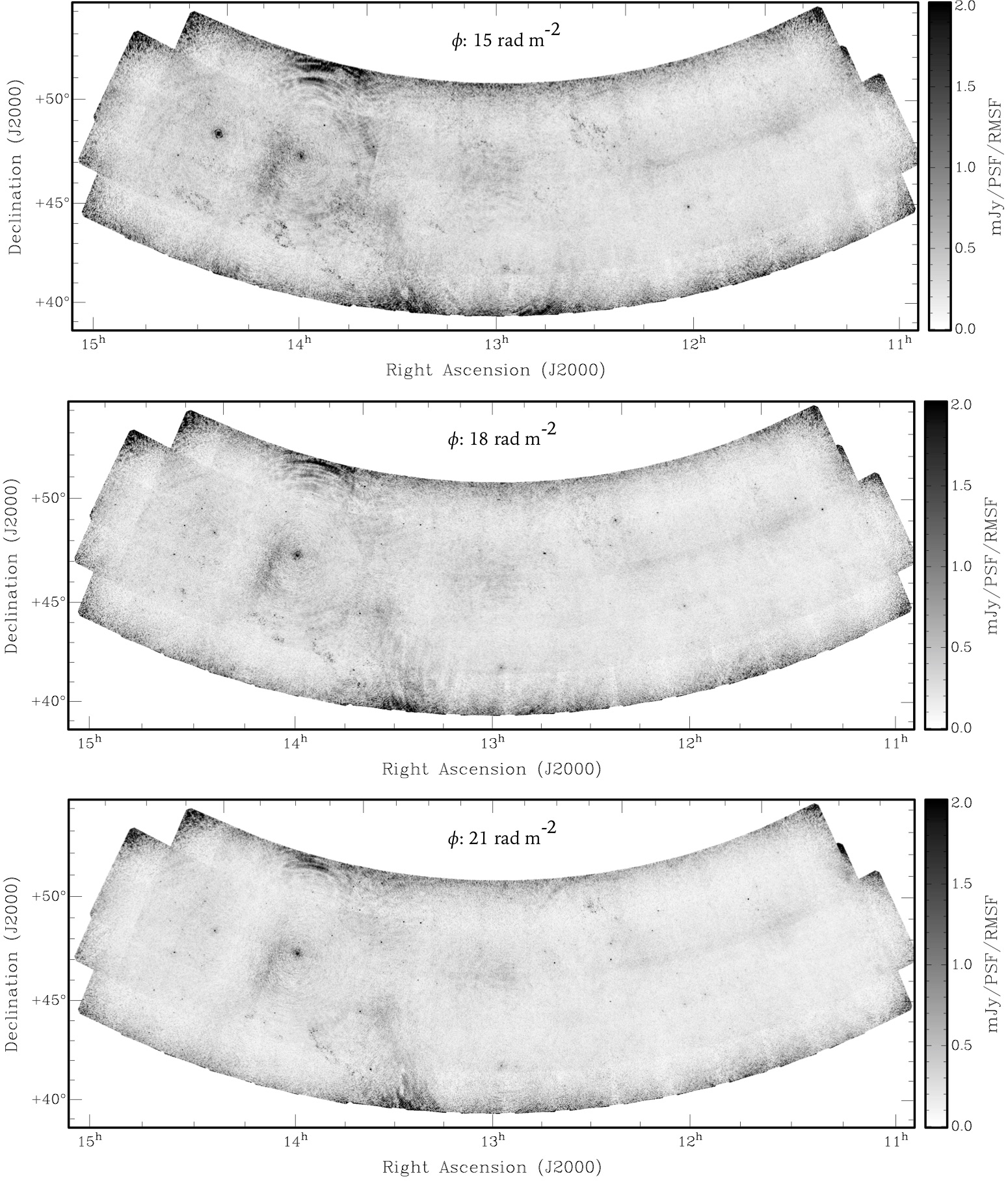} 
   \caption{Selected slices from the Faraday depth cube mosaic, showing the polarized intensity between 15 and 21 \radu. The faint, straight filamentary features slowly converge and disappear. In the top panel, a very bright polarized source appears to the left of 3C295. Some emission appears to be present at the bottom of the image, just left of center, but it is not clear if this is real or caused by enhanced noise at the edge of the mosaic.}
   \label{fig:slices6}
\end{figure*}

\begin{figure*}[t]
   \centering
   \includegraphics[width=\textwidth,height=0.9\textheight,keepaspectratio]{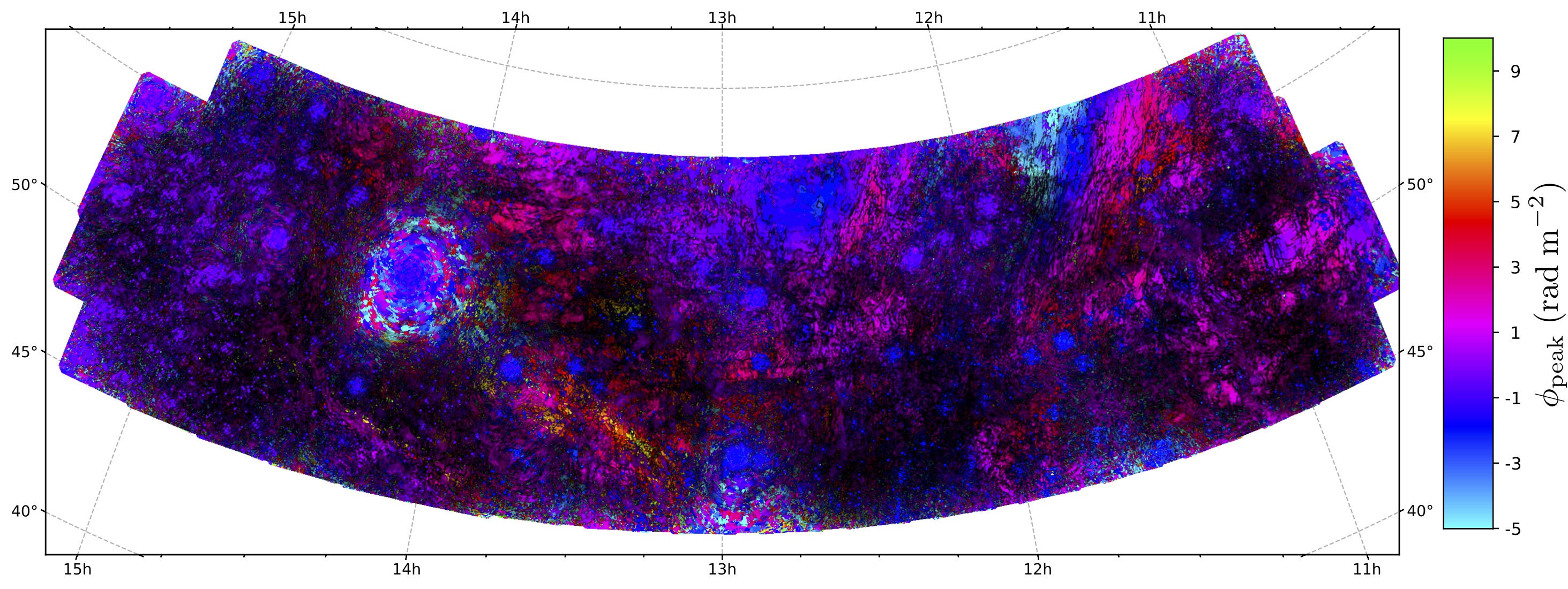} 
   \caption{Collapsed view of the full mosaic, where the polarized intensity and Faraday depth of the brightest feature in the Faraday depth spectrum were used to determine the brightness and color respectively. Instrumental leakage from bright point sources, particularly 3C295, dominate many areas with features at Faraday depths around -2 \radu, but diffuse emission can be seen through most of the mosaic.}
   \label{fig:color_full}
\end{figure*}

\begin{figure*}[htb]
   \centering
   \includegraphics[width=\textwidth,height=0.9\textheight,keepaspectratio]{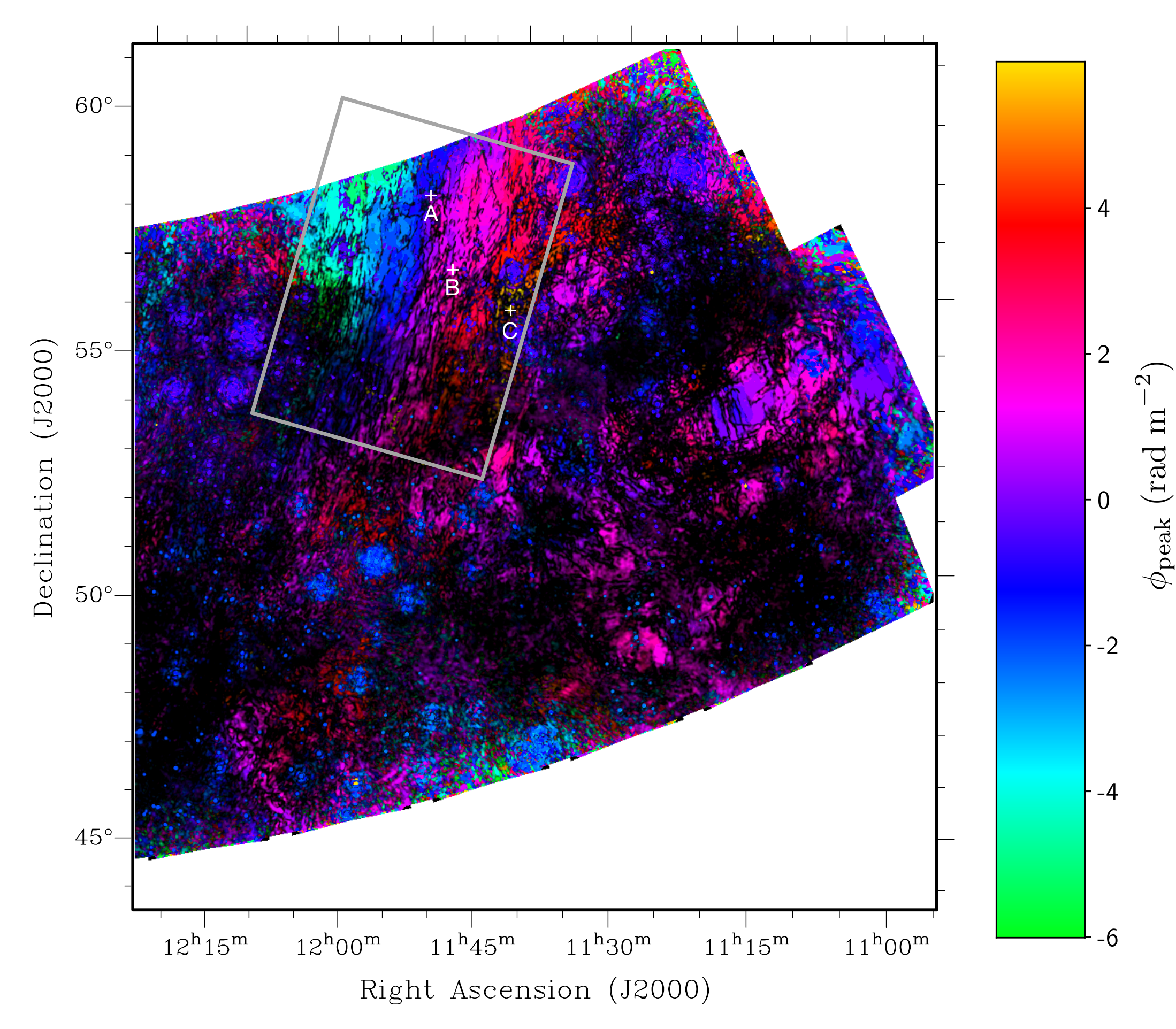} 
   \caption{Collapsed view of the west region of the mosaic, where the polarized intensity and Faraday depth of the brightest feature in the Faraday depth spectrum were used to determine the brightness and color respectively. The features in the north west and north center, marked by the grey box and described in Sect.~\ref{sec:northwest}, can be clearly seen as sheets of polarized emission with Faraday rotation gradients, while the feature in the southwest shows much more patchy structure in polarized intensity. The instrumental leakage causes bright Stokes $I$ sources to appear as small blue circular features. The three crosses mark the locations of the spectra shown in Fig.~\ref{fig:spectra}.}
   \label{fig:color_right}
\end{figure*}

\begin{figure*}[htb]
   \centering
   \includegraphics[width=\textwidth,height=0.9\textheight,keepaspectratio]{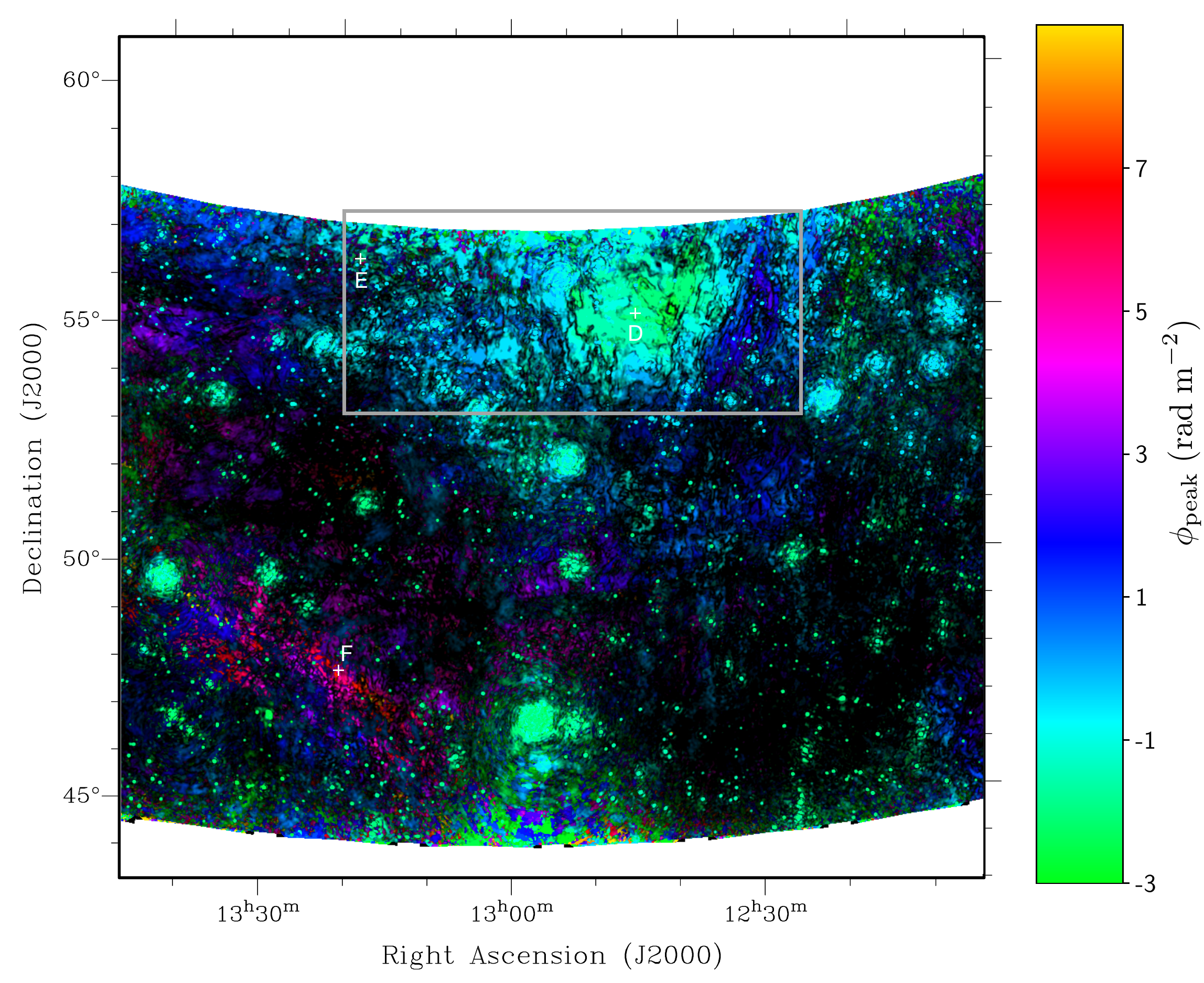} 
   \caption{As Fig.~\ref{fig:color_right}, but for the central region of the mosaic described in Sect.~\ref{sec:central_sheet}. The brightest region in the top, right of center and marked by the grey box, shows a minimum in Faraday depth of $-3$ \radu, with increasing Faraday depth in the surrounding regions. The three crosses mark the locations of the spectra shown in Fig.~\ref{fig:spectra}.}
   \label{fig:color_center}
\end{figure*}

\begin{figure*}[htb]
   \centering
   \includegraphics[width=\textwidth,height=0.9\textheight,keepaspectratio]{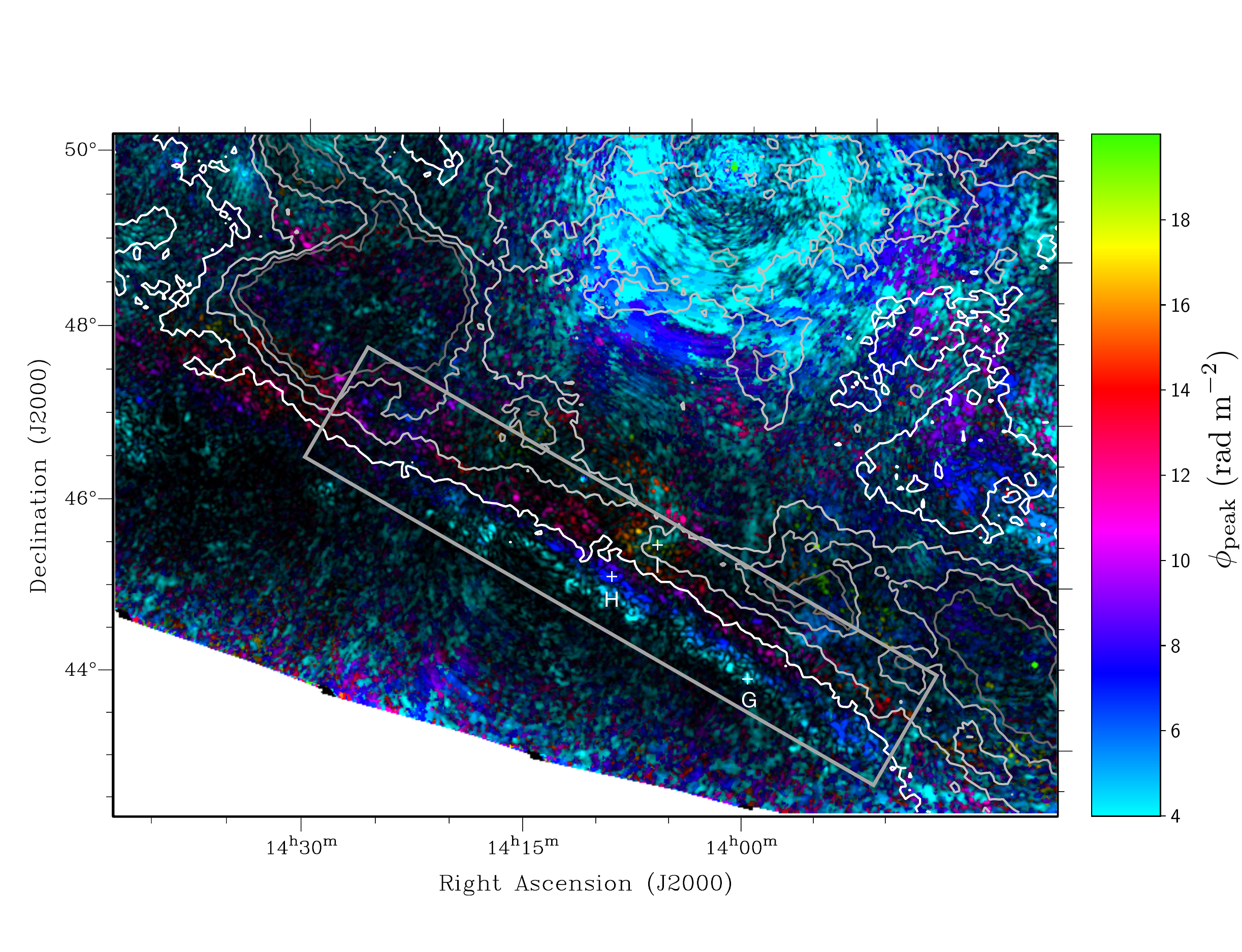} 
   \caption{As Fig.~\ref{fig:color_right}, focusing on the region around the southeast `filament' described in Sect.~\ref{sec:southeast_filament} and marked by the grey box. The contours are the \ion{H}{i} intensity integrated over the velocity range $-46$ to $-40$ km s\inv\ and were selected to highlight the \ion{H}{i} filament present in this region; the levels are 3, 6, 9, and 12 K km s\inv\, transitioning from white to dark grey. The smooth gradient in the Faraday depth is located below the edge of the \ion{H}{i} filament, while inside the filament the Faraday depth behavior is more complex. Leakage from the bright source 3C295 dominate the top of the image.  The three crosses mark the locations of the spectra shown in Fig.~\ref{fig:spectra}.}
   \label{fig:color_left}
\end{figure*}

\begin{figure}[htb]
   \centering
   \includegraphics[width=\columnwidth,height=0.9\textheight,keepaspectratio]{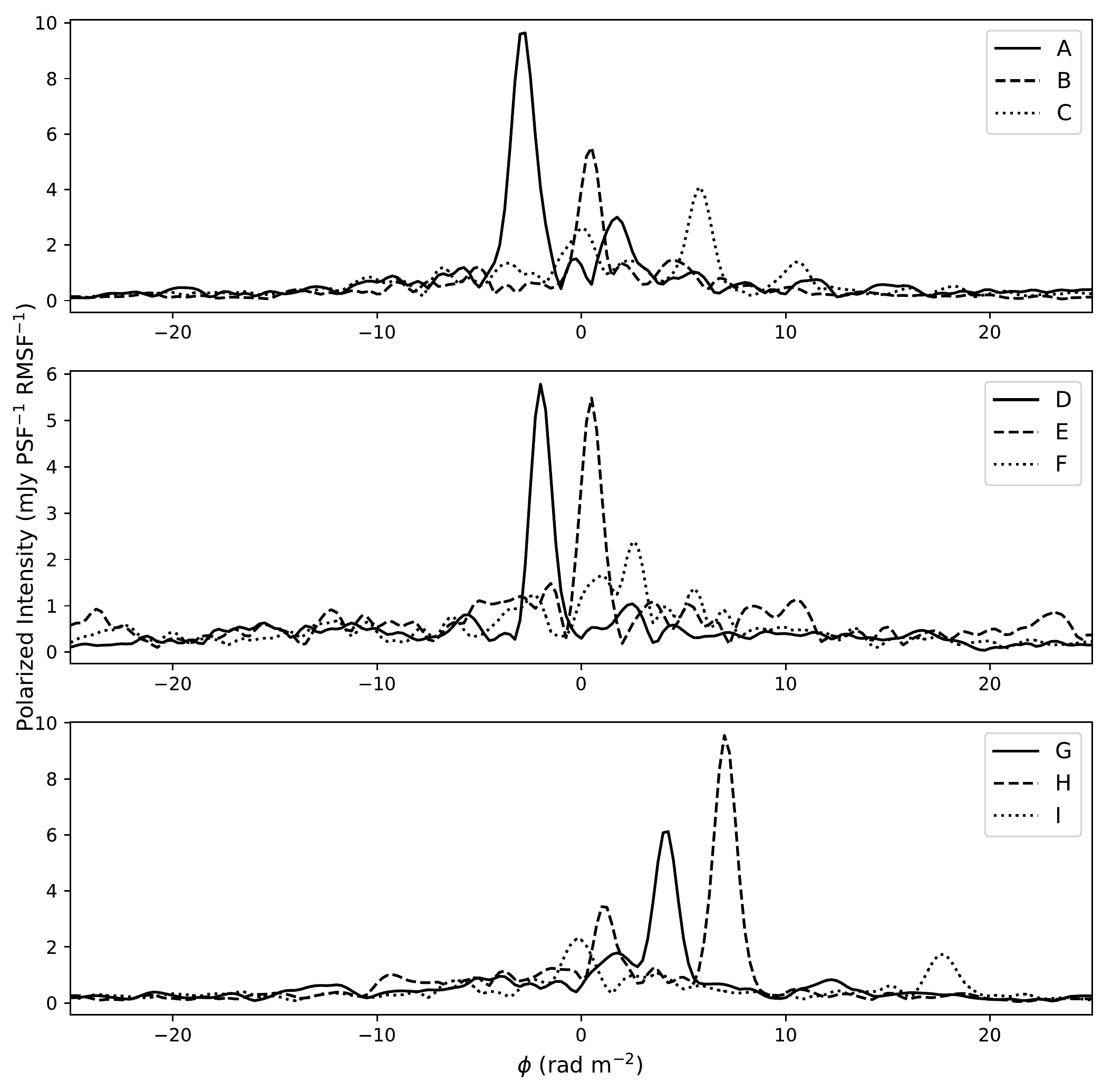} 
   \caption{Selected Faraday depth spectra from the three regions described in Sect.~\ref{sec:diffuse}, covering points in the northwest gradient (top), the central sheet (middle), and southeast filament (bottom) regions. The locations corresponding to each spectrum are shown in Figures \ref{fig:color_right} through \ref{fig:color_left}. }
   \label{fig:spectra}
\end{figure}

The combined Faraday depth cube resulting from the processing described previously covers a total area of sky of 568 square degrees, with a typical noise level between 50--100 $\muup$Jy PSF\inv\ RMSF\inv, with higher noise at the edges and near very bright Stokes $I$ sources. Figures \ref{fig:slices1} through \ref{fig:slices6} show selected slices from the cube. These images give the polarized intensity in mJy PSF\inv\ RMSF\inv; the conversion to brightness temperature units is 0.8 K (mJy PSF\inv)\inv. Figures \ref{fig:color_full} through \ref{fig:color_left} show collapsed versions of sections of the cube where, per pixel, the highest polarized intensity was located and the polarized intensity and Faraday depth of the peak were used to determine the brightness and color, respectively. Figure~\ref{fig:color_full} shows the full mosaic, while Figures~\ref{fig:color_right} to \ref{fig:color_left} focus on the specific diffuse polarized features described in Sections~\ref{sec:northwest} to \ref{sec:southeast_filament}. Figure~\ref{fig:spectra} shows some typical Faraday depth spectra from each region.
In the cube we see many diffuse polarized emission features, polarized point sources, and instrumental polarization leakage, which are discussed in detail below.

{\FloatBarrier}

\begin{figure}[ht!]
   \includegraphics[width=0.9\columnwidth,height=0.9\textheight,keepaspectratio]{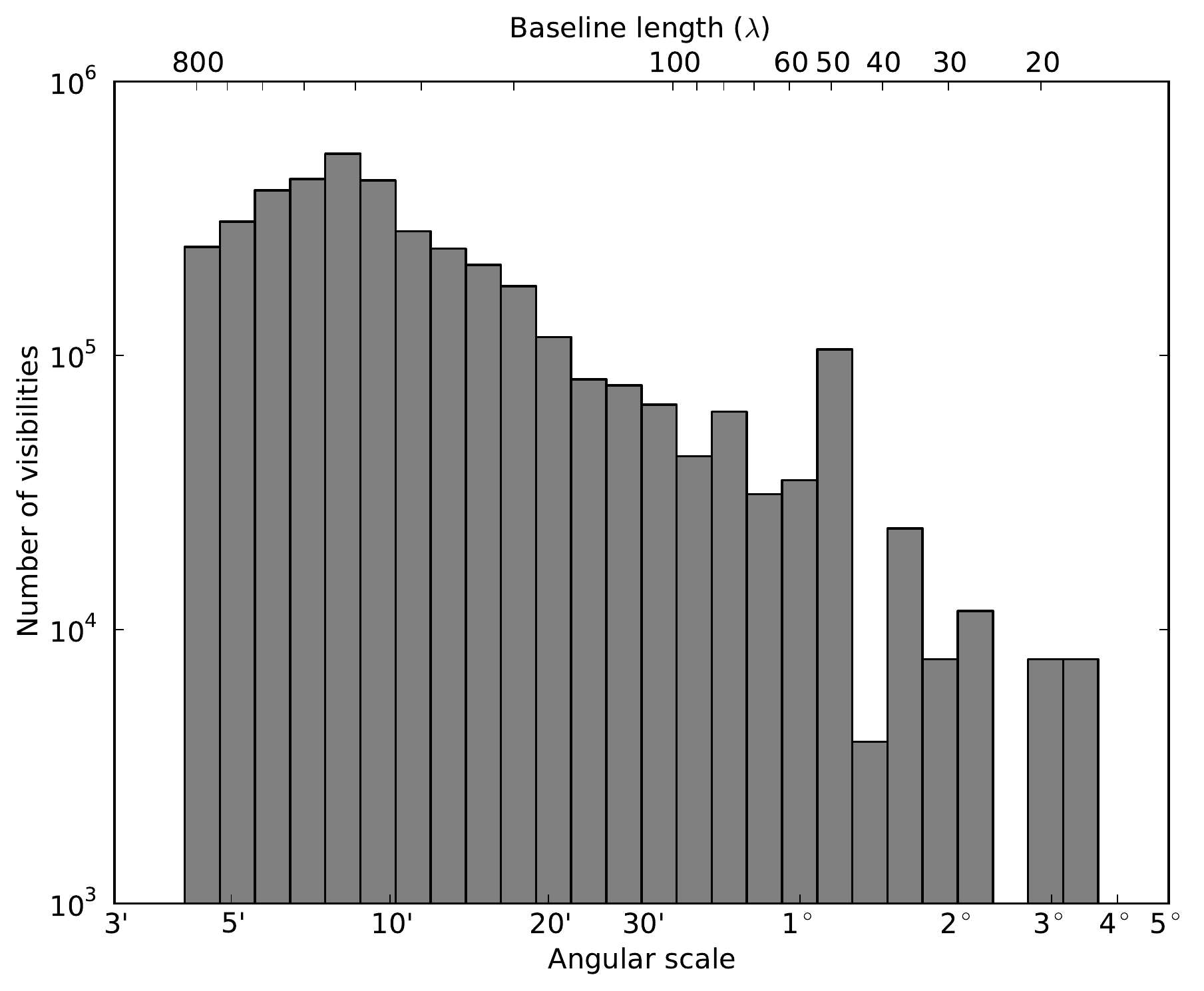} 
   \caption{ Distribution of baseline lengths in a typical LOTSS observation (after applying an upper limit of 800$\lambda$), and corresponding angular scales. The shortest baselines set the largest angular scale at approximately 3.8\degr, while the resolution of 4\farcm3 is set by the imposed limit in baseline length.}
   \label{fig:baselines}
\end{figure}

While this mosaic covers a large area of sky and allows large features to be followed across several pointings, it is not sensitive to very large polarized features. The shortest baselines present in the observations are approximately 15$\lambda$ in (projected) length, corresponding to an angular scale of about 3.8\degr; uniform polarization features larger than this scale will be missed and this should be considered when interpreting the mosaic. The distribution of baselines is shown in Figure \ref{fig:baselines}. However, at such low frequencies this may not be as significant a problem as at higher frequencies, due to Faraday rotation gradients driving emission to smaller angular scales \citep{Schnitzeler2009}. We can observe polarization features with a change in polarization angle greater than $\pi$ radians over a 3.8\degr\ angular scale; if we assume that all of this change is due to Faraday rotation fluctuations, we can calculate the amplitude of Faraday rotation variation necessary to cause this polarization change. For a typical frequency of 150 MHz, this corresponds to a Faraday depth change of 0.8 \radu\ per 3.8\degr\, or 0.21 \radu\ deg\inv. Polarization features with greater variation in Faraday depth than this value should be detected by our observations.

\subsection{Instrumental polarization leakage}\label{sec:mosaic_leakage}
The instrumental polarization leakage for LOFAR, which causes emission from Stokes $I$ to incorrectly appear in the other Stokes parameters, has the convenient property of being effectively independent of frequency. The result of this is that after RM synthesis, the leakage should appear only at Faraday depth 0 \radu, which would allow astrophysical emission at other Faraday depths to be identified. However, the ionospheric Faraday rotation correction applies a frequency-dependent polarization angle rotation to the data, which shifts the leakage to the opposite Faraday depth as the correction (e.g., if the ionospheric Faraday rotation is +2 \radu, the leakage will appear at $-2$ \radu \ after the correction). Since the ionospheric correction varies with time, this also has the result of causing partial depolarization of the leakage, as the post-correction instrumental leakage will have different polarization angles at different times.

The net effect of the instrumental polarization leakage is to cause all Stokes $I$ sources to appear in the Faraday depth cubes between approximately $-2$ and 0 \radu \ (as the ionospheric correction was typically between 0 and 2 \radu). Since the different observations had different ionospheric conditions, the leakage sources appear at slightly different Faraday depths at different locations in the mosaic, and some regions also appear to have stronger or weaker leakage depending on how much depolarization the time-variability produced.

In addition there is significant off-source polarization leakage, which is the dominant cause of the strong artifacts appearing widely distributed in Faraday depth around very bright sources such as 3C295 ($\alpha = $14$^\mathrm{h}$11$^\mathrm{m}$20.59$^\mathrm{s}$, $\delta = $+52\degr12\arcmin9.6\arcsec, 84 Jy in Stokes $I$ at 178 MHz; \citealt{Spinrad1985}). The spurious polarization caused by the polarization leakage does not appear just at the location of the source, but is convolved with the PSF, just as the Stokes $I$ and real polarized emission is. Since the locations and intensities of the sidelobes of the PSF vary with frequency, pixels near sources can pass in and out of sidelobes as a function of frequency. The result is a very complicated frequency-dependent spurious polarization signal that, when RM synthesis is applied, results in structure broadly distributed in Faraday depth. Thus, while the on-source polarization leakage is usually confined to a single Faraday depth (determined by the ionosphere), the off-source leakage is broadly distributed in Faraday depth and covers larger areas around brighter sources.

In the HETDEX region mosaic, the off-source leakage of the two brightest sources occupies a large portion of the field: 3C295 (84 Jy) is in the mid-east (left) part of the mosaic, and 3C280 ($\alpha = $12$^\mathrm{h}$56$^\mathrm{m}$59$^\mathrm{s}$, $\delta = $+47\degr18\arcmin48\arcsec, 24 Jy at 178 MHz; \citealt{Spinrad1985}) is in the bottom center of the mosaic. As a result the region around 3C295 is unusable for polarization analysis with these data.

\subsection{Diffuse emission}\label{sec:diffuse}
Diffuse polarized emission is seen through much of the mosaic, with polarized intensity levels ranging from 6--8 mJy PSF\inv \ RMSF\inv \ for the brighter features to 1 mJy PSF\inv \ RMSF\inv \ for the faintest identified features. The presence of polarization leakage has made quantitative analysis of the diffuse emission difficult, as it is not clear how to separate the diffuse emission from the off-source leakage. We have divided the diffuse emission into three regions, based on their morphologies and positions, and discuss each in turn below.

\subsubsection{Bright northwest gradient and southwest patchy emission}\label{sec:northwest}
This region, shown in Fig.~\ref{fig:color_right}, occupies the field west (right) of 12$^\mathrm{h}$ right ascension, and is dominated by two bright features that appear connected. The most striking feature is a gradient in Faraday depth that occurs in the northwest part of the field, from approximately 11$^\mathrm{h}$ to 12$^\mathrm{h}$ in right ascension and +53$^\circ$ to +57$^\circ$30$'$ in declination.\footnote{When referencing coordinates to locations in the images, it is important to recall that this is not a rectilinear projection; the projection of equatorial coordinates into our images is shown in Fig.~\ref{fig:pointing_map}.}
This emission feature appears at Faraday depths as low as $-7$ \radu\ (Fig.~\ref{fig:slices1}, top panel) and can be traced as a single continuous sheet in the cube up to +7 \radu\ (Fig.~\ref{fig:slices5}, top panel), with typical brightness of 6--8 mJy PSF\inv\ RMSF\inv. Between $-4$ \radu \ and +3 \radu \ (Fig.~\ref{fig:slices1}, bottom panel, to Fig.~\ref{fig:slices4}, top panel), this feature has the appearance of a `traveling filament', caused by a very linear gradient in Faraday depth. We measured locations along the `filament' at $-3.5$ \radu \ and at +3 \radu, and found that the separation between corresponding points was approximately 2.5$^\circ$, resulting in a Faraday depth gradient of 2.6 \radu \ deg\inv.

At Faraday depths around $-1$ \radu, additional emission emerges from many locations in the west-most third of the field and expands with increasing Faraday depth. While this emission looks patchy and disconnected in individual slices (particularly at 0, +1, and +5 \radu) and in Fig. \ref{fig:color_right}, careful inspection of the full cube shows that the emission is a continuous sheet `wrinkled' in Faraday depth, with many local minima and maxima in Faraday depth surrounded by a complex web of emission at intermediate Faraday depths. This is best seen at +3 \radu \ (Fig.~\ref{fig:slices4}, top panel) where many `loops' of emission can be seen (surrounding the locations of local Faraday depth maxima or minima); this slice also best shows the extent of the emission, which extends the full height of the field.

Near the region occupied by the linear gradient, a second emission feature is also present at higher Faraday depths. It can be seen at +3 \radu \ at the same location occupied by the first feature at -5 \radu\ (as a result, it cannot be seen in Fig.~\ref{fig:color_right}, which only shows the brightest Faraday-depth peak found in each pixel). It is seen over a much smaller area than the first feature, but this may be due to the much lower brightness of this feature, 1--2 mJy PSF\inv \ RMSF\inv. It can be seen at Faraday depths up to +14 \radu, and while it has some morphological similarities to the first feature (gradients and local maxima at similar locations), it also has some significant differences (a notable absence of emission in certain regions that are bright in the first feature).

\subsubsection{Central sheet}\label{sec:central_sheet}
The region between 12$^\mathrm{h}$ and 14$^\mathrm{h}$, shown in Fig.~\ref{fig:color_center}, is also dominated by a bright feature. This feature appears to be a continuous sheet of emission distributed in Faraday depth between $-3.5$ and +12 \radu, with several local maxima and minima in Faraday depth at various locations and filamentary-looking emission. The typical brightness is 3--4 mJy PSF\inv \ RMSF\inv. The brightest region occurs at the northern edge of the field, where emission appears at $-3.5$ \radu\ and spreads outwards in all directions with increasing Faraday depth, with a typical brightness of 6--8 mJy PSF\inv \ RMSF\inv. A second bright spot occurs to the southeast, around (13$^\mathrm{h}$15\arcmin,47\degr), with Faraday depths of 5--7 \radu.

\subsubsection{Southeast `filament'}\label{sec:southeast_filament}
The third region, shown in Fig.~\ref{fig:color_left} is between 14$^\mathrm{h}$ and 15$^\mathrm{h}$ right ascension, and between +45\degr\ and +50\degr\ declination, and contains another continuous feature with a very linear Faraday depth gradient. The feature appears as a very long thin filament in Faraday depth slices from +1 \radu \ to +11 \radu, which in some slices can be seen to extend over 12\degr\ in length. As with the other features, this can be interpreted as a continuous sheet of polarized emission. Around +9 \radu, additional emission can be seen further north, and as Faraday depth increases these merge into the filaments toward several local maxima at Faraday depths between +19 and +25 \radu. The typical brightness of this feature is 1--2 mJy PSF\inv \ RMSF\inv. The change in position of the filament was measured between +3 \radu \ and +11 \radu, and was found to be 0.6\degr, indicating a Faraday depth gradient of 13 \radu \ deg\inv.

\subsection{Point sources}
In addition to diffuse Galactic polarized emission, polarized emission is also observed from some of the unresolved point sources. The identification, measurement, and analysis of these sources is done in a parallel paper \citep{VanEck2018a}, and will not be discussed here. The majority of sources are found at Faraday depths between +10 and +25 \radu, higher than most of the diffuse emission. This implies that the observed diffuse emission is from only a portion of the line of sight through the Galaxy, with the more distant parts contributing additional Faraday rotation to the extragalactic sources.

\section{Comparison with other tracers}\label{sec:tracers}
Previously published studies of Faraday depth cubes have looked at relating the observed diffuse polarization to tracers of other ISM components. \citet{Zaroubi15} related Faraday depth structure observed in one LOFAR field to high-frequency (353~GHz) polarization maps, which trace dust emission, from the {\it Planck} mission, and found that the observed filaments in Faraday depth followed the magnetic field orientation inferred from the {\it Planck} polarization. \citet{VanEck17} suggested that neutral clouds could act as sources of Faraday-thin polarized emission (which would not be strongly depolarized at low frequencies), and associated two observed polarization features with neutral clouds in the local ISM.

We investigated several ISM tracers for features corresponding to those we see in our Faraday depth cube. For each tracer, we regridded the data onto the same projection as our field, for ease of comparison. These tracers and their sources are:
\begin{itemize}
\item H$\alpha$ (integrated over velocity), from \citet{Finkbeiner03}, with a resolution of 1\degr.
\item 408 MHz radio continuum, from \citet{Remazeilles15,Haslam1982}, with a resolution of 56\arcmin.
\item {\it Planck} thermal dust emission, with a resolution of 60\arcmin, at a reference frequency of 545 GHz; all {\it Planck} maps taken from the Commander component separation in \citet{Planck2015_components}.
\item {\it Planck} thermal dust polarization, with a resolution of 10\arcmin, at a reference frequency of 353 GHz.
\item {\it Planck} synchrotron emission, with a resolution of 60\arcmin, at the reference frequency of 408 MHz.
\item  {\it Planck} synchrotron polarization, with a resolution of 40\arcmin, at the reference frequency of 30 GHz.
\item  {\it Planck} CO(1-0), with a resolution of 60\arcmin.
\item  {\it Planck} free-free emission, with a resolution of 60\arcmin.
\item integrated \ion{H}{i} line emission from the Effelsberg-Bonn HI survey \citep[EBHIS,][]{Winkel16}, with a resolution of 11\arcmin.
\item dust extinction, calculated from \citet{Green15} as described below, with a resolution of 42\arcmin.
\item 1.4 GHz polarized intensity, from \citet{Wolleben2006}, with a resolution of 36\arcmin.
\end{itemize}
The maps of each of these tracers are shown in Figures \ref{fig:tracers1} and \ref{fig:tracers2}. The {\it Planck} polarization maps were converted from Stokes $Q$ and $U$ to polarized intensity. The extinction maps were made by using the MWDUST package \citep{Bovy16} to get the optical extinction as a function of distance, taking the numerical derivative with respect to distance (which should serve as a proxy for dust density). This produced a 3D data cube of extinction per unit distance, which was then integrated over selected distance ranges to produce extinction maps for different distances. The distance ranges were chosen to be 0--75 pc, 75--250 pc, and 250--1000 pc. The boundary at 75 pc was chosen as the \citet{Green15} extinction model is often poorly constrained for distances less than this; the boundary at 1000 pc was chosen as no significant extinction contribution was seen beyond this distance; the boundary at 250 pc was chosen as the extinction showed different spatial structure on both sides of this boundary (i.e., the morphological differences between the two were maximized by placing the boundary at 250 pc). The extinction maps were observed to have a large scatter between adjacent pixels, so they were smoothed using a Gaussian kernel with a FWHM of 42\arcmin\ ($\sigma$=18\arcmin) to produce the maps shown in Fig. \ref{fig:tracers2}.

\begin{figure*}[p]
   \centering
   \includegraphics[width=0.95\textheight,height=\textwidth,keepaspectratio,angle=90]{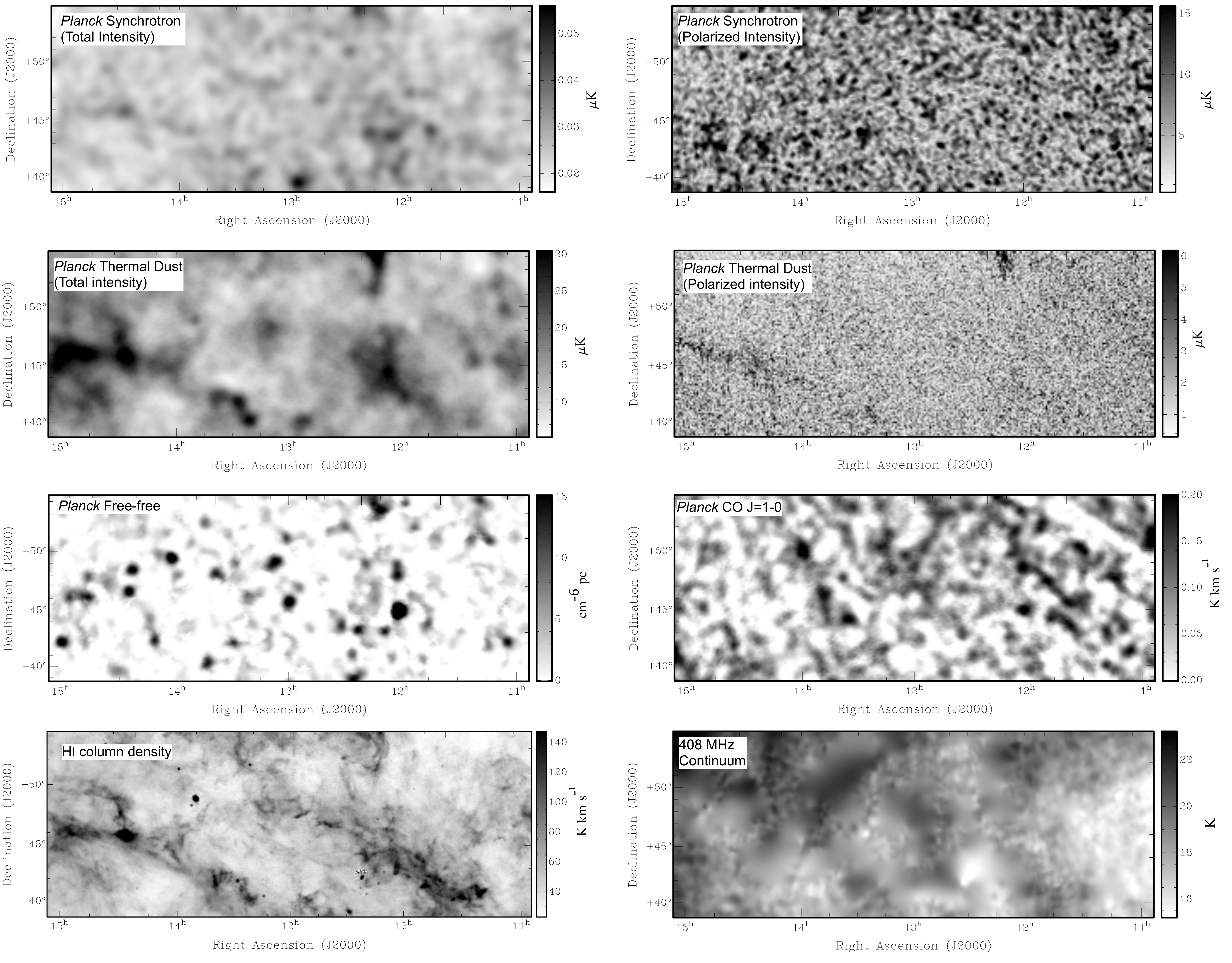} 
   \caption{Selected ISM tracers, in the same coordinates and projection as the previous images. The \ion{H}{i} filament in the southeast region of the field is well correlated, but slightly offset in position, with the steep linear gradient in Faraday depth described in Sect.~\ref{sec:southeast_filament}; this is shown in more detail in Fig.~\ref{fig:color_left}.}
   \label{fig:tracers1}
\end{figure*}

\begin{figure*}[p]
   \centering
   \includegraphics[width=0.95\textheight,height=\textwidth,keepaspectratio,angle=90]{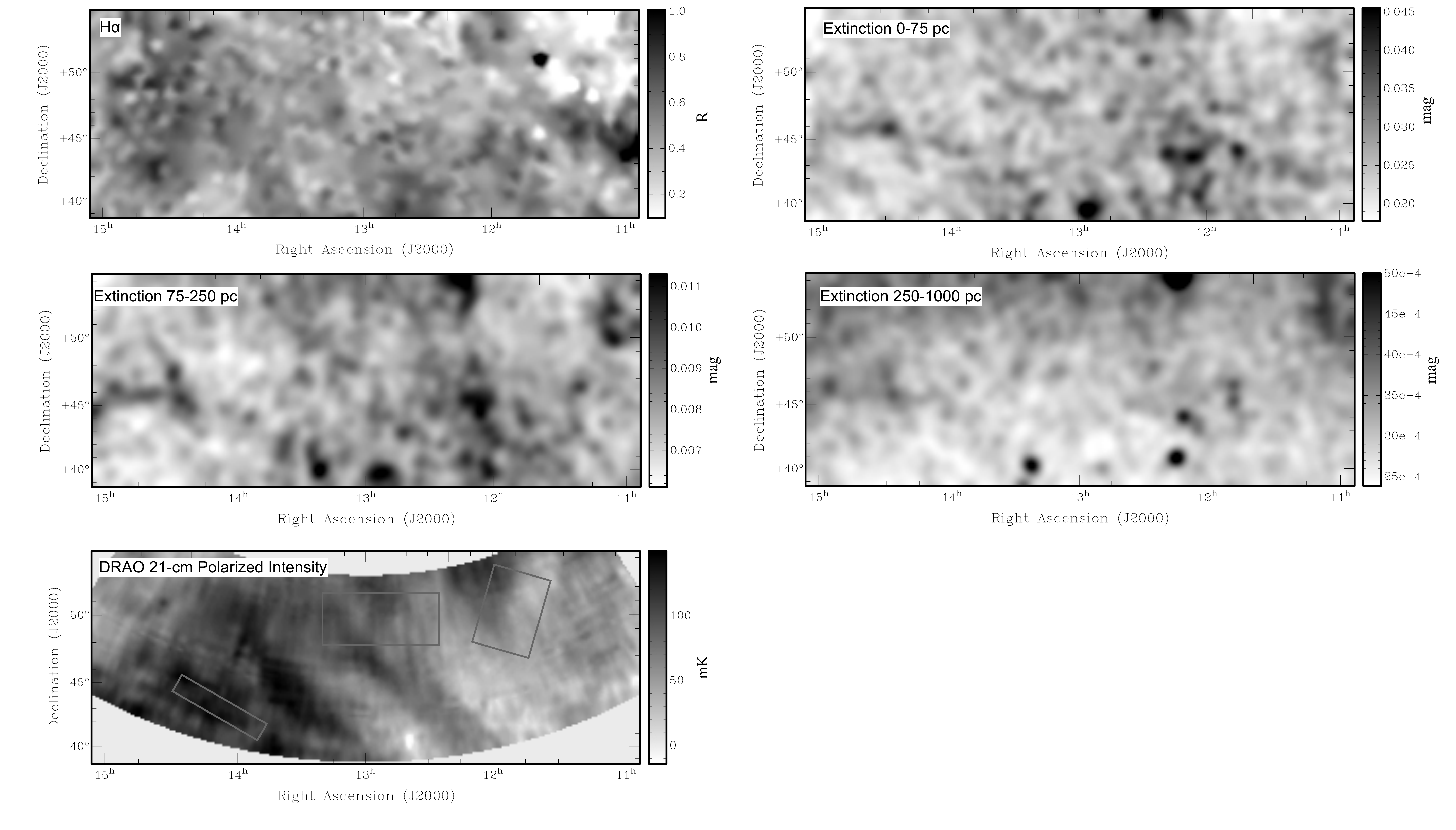} 
   \caption{Continuation of figure \ref{fig:tracers1}, with additional tracers. The DRAO 21-cm polarized intensity shows stronger emission in several regions where we observe bright features and maxima/minima in Faraday depth in our data. To guide the eye, the grey boxes from Figs.~\ref{fig:color_right}-\ref{fig:color_left} have been overlaid.}
   \label{fig:tracers2}
\end{figure*}

=

The synchrotron emission and polarization, 408~MHz radio continuum, H$\alpha$, free-free emission, and CO emission maps are all largely featureless, so no comparisons with the low-frequency radio polarization were possible. The \ion{H}{i} column density, thermal dust emission and polarization, and extinction between 75--250 pc all show similar structure, and the location of this structure does not appear to strongly correlate with the position of the low-frequency polarized emission.

However, there is a very interesting relationship between the southeast Faraday depth gradient feature and the \ion{H}{i} filament in the southeast part of our field. Figure \ref{fig:color_left} shows this region of the field, comparing our Faraday depth cube with contours made by integrating the \ion{H}{i} data over the velocity range of the filament ($-46$ to $-40$ km s\inv). The \ion{H}{i} filament has a sharp boundary with minimal emission (at any velocity) south of the lowest contour. The linear gradient in Faraday depth occurs south of the \ion{H}{i} filament, and at the location of the filament the Faraday depth structure is more complex, with several local maxima, and the polarized intensity is more patchy. This transition from linear gradient to more complicated structure is very closely aligned with the boundary of the filament, including the slight bend around $\alpha = $14$^\mathrm{h}$30$^\mathrm{m}$, $\delta = $+47\degr20\arcmin, which is strong evidence that this is not coincidental positioning. We discuss interpretations of this relationship in Sect.~\ref{sec:HIdiscussion}. We note that while the southeast gradient shows this clear alignment with an \ion{H}{i} filament, this is not the case for the northwest gradient, which does also occur in a region of low \ion{H}{i} emission but has no \ion{H}{i} features in the area where the gradient transitions to a more complex morphology.

The comparison of our 150 MHz polarization data to the 1.4 GHz polarization data from \citet{Wolleben2006} is interesting, but complicated by a number of factors. First is that we are comparing a Faraday depth cube, which can separate out multiple Faraday depth components, to a frequency-averaged polarization map, which will show a combination of all Faraday depth components present. Second, and related to the first, is that our low-frequency data will not be sensitive to Faraday-thick components while the higher frequency data may include emission from such components (albeit possibly partially depolarized). Third, the 1.4 GHz data is at a much lower resolution, 36\arcmin, so it will likely be more strongly affected by beam depolarization.

With those caveats, there are some interesting relationships between the 1.4 GHz polarized intensity and some of the features in our Faraday depth cube. Several of the brighter regions in the 1.4 GHz data correspond to brighter regions in the Faraday depth cube: the brighter region in the northwest region, around (11$^\mathrm{h}$30\arcmin, +57\degr), overlaps with the bright, low-Faraday depth region in the northwest gradient (the green region in Fig.~\ref{fig:color_right}); the bright region in the north-center, around (13$^\mathrm{h}$, +57\degr), overlaps the brightest part of the central sheet (the bright green region at the top of Fig.~\ref{fig:color_center}); and the bright region slightly east and south of the center, around (13$^\mathrm{h}$30\arcmin, +49\degr), overlaps with the second bright region in the central sheet (red region in the lower left of Fig.~\ref{fig:color_center}). It is particularly noteworthy that each of these regions corresponds to a maximum or a minimum in Faraday depth, and that the regions with intermediate, more spatially variable Faraday depths show weaker polarized intensity at 1.4 GHz. This is not a perfect correlation: the bright 1.4 GHz region in the southeast region of the field, around (14$^\mathrm{h}$30\arcmin, +47\degr), overlaps with the steep Faraday depth gradient (the `filament' in Fig.~\ref{fig:color_left}).

One possible explanation is that this might be depolarization caused by the Faraday depth gradients. However, this is not supported by the results of Appendix \ref{sec:depolarization}, as the gradient of 2.5 \radu \ deg\inv\ would produce only a 4 degree change in polarization angle across the 1.4 GHz beam, so minimal depolarization by the gradient is expected. Based on this, we rule out this explanation.

Finding an explanation becomes more difficult when the polarized intensities are compared. We estimate the typical total-intensity spectral index in this region is -2.6 (using the 150 MHz data from \citealt{Landecker1970} and the 1420 MHz data from \citealt{Reich82}), so under the simplest assumption of a constant polarized fraction (perfectly Faraday-thin emission) the expected polarized intensity at 150 MHz can be estimated. In the areas of faintest 1.4 GHz polarization, the polarized brightness temperature at 1.4 GHz is typically 20--40 mK, which would be equivalent to 7--14 K at 150 MHz. Correspondingly, the brighter areas with polarized brightness temperatures of 100--150 mK would be equivalent to 35--50 K. In the fainter regions, where the northwest gradient is, the typical polarized intensity we observe at 150 MHz is about 3--5 K; correspondingly in the brightest regions of the central sheet we observe maximal intensities of about 8 K. 

These values imply we are only detecting roughly 40\% of the polarized intensity in the faint regions and about 20\% in the brighter regions. The two most straightforward explanations for this are that the `missing' polarized intensity is either in Faraday-thick components and correspondingly depolarized at low-frequencies, or exists on the very large angular scales not observed by LOFAR. However, it's not clear why either of these effects would cause a higher degree of missing emission in regions of higher polarized intensity.

In retrospect, it may not be surprising that many of the ISM tracers show very little signal or structure. The HETDEX field was originally selected for high-redshift cosmology observations in the optical, and was chosen for minimal contamination from Galactic foregrounds. As a result, many of the tracers we investigated were barely detected in this region. In this respect, the choice of initial region for LOTSS is somewhat unfortunate for Galactic foreground science, but it also suggests that future investigations using observations in other regions, particularly at lower Galactic latitudes, will probably have more visible ISM structure against which to compare the Faraday tomography observations.

\section{Interpretation of diffuse emission}\label{sec:interpretation}
The lack of correlations between the polarized emission and the ISM tracers provides constraints on the source of the polarized emission. Below we consider separately the source of the Faraday thin polarized emission, and the Faraday rotation of that emission.

\subsection{Origin of Faraday thin emission}\label{sec:caustics}
Due to the frequency coverage of our observations, we are insensitive to polarized features with Faraday thickness greater than about 1.0 \radu. This places strong constraints on the ISM structures that we can observe; \citet{VanEck17} discussed which ISM conditions could cause Faraday-thin features. Here we consider these conditions in relation to the polarization features were observed.

A localized enhancement in the magnetic field perpendicular to the line of sight, such as in a shock, could produce a Faraday-thin feature, but would also produce a synchrotron excess which should be seen in total intensity. An enhancement in the degree of order in the magnetic field could also produce such features. There are also no obvious sources of such enhancements, such as supernova remnants, in this field, so both of these possibilities seem unlikely.

Strong localized enhancements or diminishments in the free electron density could also produce Faraday thin features. Enhancements would be associated with objects like shocks or \ion{H}{ii} regions, neither of which are observed in this field. Typical locations of low free electron density would be neutral clouds (associated with the warm neutral or cold neutral ISM phases) or regions of the hot ionized ISM phase. The tracers of neutral material (\ion{H}{i}, thermal dust, extinction) do not show any correlation to the observed polarized emission, and the local ISM models of \citet{Lallement14} do not show any clouds within the nearest few 100 pc in this direction. The Local Bubble is a volume of hot ionized medium (HIM) surrounding the Sun, and may (depending on the path length in this direction) have internal Faraday rotation of less than 1 \radu\ which would result in a Faraday thin feature \citep{VanEck17}. However, this feature would not have any foreground to produce Faraday rotation, so the feature would appear at 0$\pm$0.5 \radu and could not be the source of the emission features seen at other Faraday depths. More distant regions of HIM could be the source of that emission, provided they are not so large as to become Faraday thick.

Another possibility is a region where the component of the magnetic field parallel to the line of sight is very small. If this occurs as a result of a reversal in the sign of the parallel magnetic field this is called a Faraday caustic \citep{Bell11}, but lines of sight where the parallel component is very small in some volume but has the same sign in front and behind this volume are also possible. Such regions are possible in any phase of the ISM. \citet{Bell11} showed that Faraday caustics can form sheets of Faraday-thin emission, covering the area of sky with the sign-reversal. While \citet{Bell11} show that Faraday caustics should have a distinctive one-sided tail in the Faraday depth spectrum, they also state that to resolve this structure requires that the ratio between the highest and lowest frequencies must be at least 1.5 (their equation 19), which is not satisfied for our observations. As a result we cannot identify this characteristic feature of Faraday caustics, but future observations with larger bandwidth may be able to do so.

Finally, these features could represent sharp `edges' in the intrinsic Faraday spectra, caused by sharp transitions in the Faraday spectrum amplitude at the edges of Faraday thick features. Such transitions would have similar causes as the Faraday-thin features described above: a sharp transition in the free electron density, the strength of the parallel component of the magnetic field, or the synchrotron emissivity. However, such sharp edges tend to be strongly depolarized \citep{VanEck17}, which in turn requires a very high intrinsic polarization to produce a detectable feature in the Faraday depth cubes.

We are not able to determine which of these possible causes are the source of the Faraday-thin features we observe. The potential signatures in the other ISM tracers of these conditions are difficult to measure due to limited sensitivity of the tracer observations, or to the fact that most of the tracer data are line-of-sight integrated rather than tomographic measurements.

\subsection{Gradients in Faraday rotation}
The Faraday depth structure seen in the diffuse emission shows a wealth of complex morphologies, such as the linear gradients and local minima/maxima previously described. A detailed analysis of all these individual features is beyond the scope of this work, but we do consider some possible causes of the Faraday depth gradients.

A gradient in Faraday depth can be caused by a gradient in the three factors that determine the Faraday depth: the free electron density, the parallel magnetic field, and the (physical) depth of the Faraday-rotating volume. A gradient in the depth of the Faraday-rotating region would naturally result if the emitting region was a sheet that was not perpendicular to the line of sight and the rotation occurred directly in front of the emitting region; if such an emitting sheet was reasonably flat the distance to the region, and the corresponding Faraday depth, could increase linearly with position. Gradients in the electron density might be expected on larger (kpc) scales, due to effects like the scale height of free electrons, and also on the scales of individual clouds (e.g., a transition layer between a neutral cloud and an ionized skin or exterior).

\begin{figure}[htbp]
   \centering
   \resizebox{\hsize}{!}{\includegraphics[width=\textwidth]{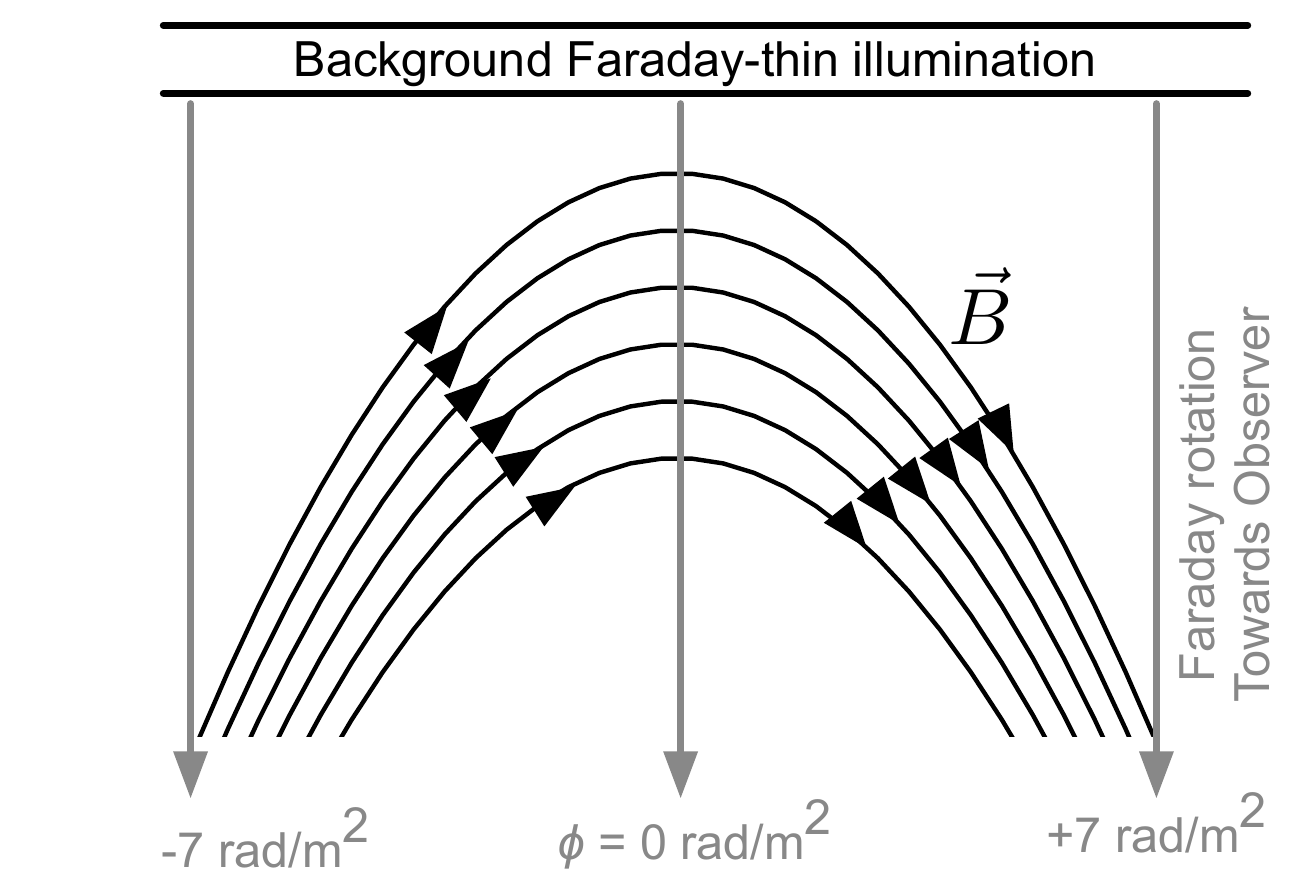}} 
   \caption{A schematic diagram of a possible source for Faraday depth gradients, where a magnetic loop (black lines) would possess a gradient in the parallel component of the magnetic field and in turn produce a gradient in Faraday depth. An (unrelated) Faraday-thin emission feature in the background would provide the observed source of polarized emission (grey lines).}
   \label{fig:magnetic_loop}
\end{figure}

The linear gradient in the northwest feature (Sect. \ref{sec:northwest}) contains both negative and positive Faraday depths. The simplest explanation for this is a gradient in the parallel component of the magnetic field that causes the sign to change from negative to positive. Alternative explanations would require two regions at different distances with oppositely directed parallel magnetic fields and an electron density or distance gradient in one of the regions. However, this would produce a Faraday caustic associated with the reversal in parallel magnetic field between these two regions and a corresponding Faraday-thin emission feature (which we do not observe). While there are many possible magnetic field configurations that would produce a gradient in the parallel component of the magnetic field, one possible configuration that would be consistent both with the observations and theoretical models \citep[as discussed in ][]{Beck96,Shukurov04} is a magnetic filament bent into a loop such as is shown in Figure \ref{fig:magnetic_loop}.

\subsection{Southeast Faraday depth gradient-HI correlation} \label{sec:HIdiscussion}
The transition from a smooth linear gradient in Faraday depth outside of the \ion{H}{i} filament, shown in Fig.~\ref{fig:color_left}, to a more complicated morphology overlapping the filament is very intriguing. Since we are able to observe smooth transitions in the Faraday depth and polarized intensity in the filament region, we conclude that the \ion{H}{i} filament is not associated with the source of the polarized emission, and must lie in the foreground with the rest of the Faraday-rotating structure. Furthermore, since the orientation of the gradient so closely aligns with that of the filament boundary, we conclude that the gradient is most likely related to, and co-distant with, the \ion{H}{i} filament.

One possible interpretation of the source of the gradient is in terms of an envelope of ionized material surrounding the (predominantly neutral) \ion{H}{i} filament. If this is the case, then we expect to see a similar gradient on the other side of the filament; unfortunately that corresponds to the region where instrumental leakage from 3C295 overwhelms the real signal. Improvements in the leakage calibration for LOFAR data should allow for future investigation of this possibility.

\section{Depolarization by Faraday depth gradients}\label{sec:gradients}
In addition to constraining possible magnetic field configurations, the Faraday depth gradients seen in this field are noteworthy in that they can be difficult to observe, requiring specific observational parameters. At higher frequencies, it is often the case that the Faraday depth resolution is so low that gradients in Faraday depth can only be measured in cases with very high signal-to-noise ratios. However, at lower frequencies the synthesized beam is typically larger than at higher frequencies, leading to stronger beam depolarization masking the polarized signal and making the gradients undetectable. Appendix \ref{sec:depolarization} describes this beam depolarization, and gives an approximate threshold for the conditions under which the depolarization becomes significant: polarization angle gradients greater than 180\degr\ over the FWHM of the beam will be strongly depolarized, and gradients shallower than this may also be significantly depolarized depending on the exact beam parameters.

We can use this threshold to determine the depolarization caused by the Faraday depth gradients in our observations. For our observations, with a FWHM of 4\farcm3 and wavelengths between 1.9 m and 2.5 m, a polarization angle gradient of 180\degr\ FWHM\inv \ would be created by a Faraday depth gradient of 7.5--13 \radu \ deg\inv. In the HETDEX field, we observed two very linear gradients in Faraday depth, with values of 2.6 \radu\ deg\inv\ (Sect.~\ref{sec:northwest}), which is well inside the predicted range of minimal depolarization, and 13 \radu \ deg\inv\ (Sect.~\ref{sec:southeast_filament}), at the upper limit of this range. Beam depolarization by the gradient may explain the lower polarized flux density of the steeper gradient compared to the other polarized features. For these gradients, the Gaussian model predicts that the shallower gradient would be detected at 87\% of its intrinsic intensity and the steeper gradient at 3\%. Our synthesized beam likely falls somewhere between the two models, implying that the measured brightness of the shallower gradient is probably accurate within approximately 10\% while the steeper gradient is probably somewhere between 1--30 times brighter than our measurement.

We can also consider the detectability of Faraday depth gradients with other low-frequency radio observations. \citet{Lenc16} observed faint diffuse emission with the MWA, at 154 MHz ($\lambda$ = 1.95 m) with a resolution of 54\arcmin\ and similar sensitivity as our observations. With the simple beam model of Appendix \ref{sec:depolarization}, the depolarization threshold corresponds to a Faraday depth gradient of 0.9 \radu \ deg\inv. However, they also applied a Gaussian ($u,v$) taper to their data, which probably has the effect of making their beam more Gaussian-like and increasing the depolarization present. These observations are probably strongly affected by beam depolarization due to Faraday depth gradients; the majority of the structure seen in the HETDEX field could not be detected with these observational parameters. Observing such features may be possible with the MWA by including the longer baselines and minimizing the effects of the Gaussian taper.

There are plans to use the data from the Canadian \ion{H}{i} Mapping Experiment (CHIME) to measure polarized foregrounds. CHIME will operate at 400--800 MHz, with a resolution of approximately 30\arcmin. The Faraday depth gradient threshold for these parameters varies from 11--45 \radu \ deg\inv \ from the bottom of the frequency band to the top, indicating that these observations should be less affected by gradient beam depolarization than our LOTSS observations.

The sensitivity of observations to steep Faraday depth gradients may be a consideration in planning future observations, as it may have strong consequences on the interpretation of such observations. If polarized features with steep gradients are present in an observed region of sky, they may strongly depolarized and thus not detected. This would lead to an underestimate in the amount of polarized flux present, and could introduce a bias into any inferred properties of the diffuse polarization and ISM magnetic field. Also, polarized features that include regions of both shallow and steep gradients would have missing regions, which could cause large connected Faraday depth structures to be interpreted as separate features.

We have not used the full resolution available with the LOTSS data, so a reprocessing of these data at higher resolution could improve the sensitivity to steeper Faraday depth gradients. At the same resolution as the initial LOTSS Stokes $I$ data products, 25\arcsec, the threshold for depolarization increases to 72--125 \radu \ deg\inv \ at the bottom and top ends of the frequency band. However, the higher resolution results in a correspondingly lower sensitivity to diffuse emission, making it difficult to select a compromise between maximizing sensitivity to diffuse emission while preventing beam depolarization due to unresolved polarization angle gradients. Higher resolution also significantly increases the computational and data storage requirements of such reprocessing, so future surveys will need to balance computational expense against sensitivity to steeper Faraday depth gradients.

\section{Conclusions}\label{sec:conclusions}
We have used 60 8-hour observations from the LOFAR Two-meter Sky Survey to perform Faraday tomography covering the HETDEX Spring field (right ascension from 10$^\mathrm{h}$30$^\mathrm{m}$ to 15$^\mathrm{h}$30$^\mathrm{m}$ and declination from 45\degr\ to 57\degr). We have produced a mosaic Faraday depth cube of this region at 4\farcm3 resolution, which shows polarized emission as a function of Faraday depth. The low-frequency nature of our data gives us a Faraday depth resolution of 1 \radu, allowing us to probe very small variations in Faraday depth. We achieve a typical sensitivity of 50--100 $\muup$Jy PSF\inv\ RMSF\inv.

In our Faraday depth cube we see diffuse polarized emission across most of the region, at Faraday depths between $-7$ \radu\ and +25 \radu. This diffuse emission mostly takes the form of `sheets', where the emission appears to be filamentary at any single Faraday depth but can be seen as a continuous feature distributed smoothly over Faraday depth. We are able to map out several of these sheets in different positions in the region. A few of these show very linear features, where emission over several square degrees has a smooth linear gradient in Faraday depth.

We compared our Faraday depth cube with several other tracers of different ISM components, and found two interesting relationships. One of the large linear Faraday depth gradients follows the edge of an \ion{H}{i} filament, with the gradient occurring just outside the observed edge of the filament with the Faraday depth decreasing with distance from the filament. We also found that the locations of maxima and minima in the Faraday depth of the features we observe tend to overlap with regions of high 21-cm polarized intensity.

Motivated by the Faraday depth gradients seen in our observations, we considered beam depolarization caused by the polarization angle gradients produced by such Faraday depth gradients. Many previous authors have considered the depolarization effects of a Gaussian PSF, which produces very strong depolarization. We have shown that an idealized PSF for an interferometer with uniform $(u,v)$ coverage causes much weaker depolarization for some cases. We expect that most realistic observations will fall between these two cases, and suggest that Faraday depth gradients may be more likely to be detected than previously expected.

We have shown that the LOTSS data are well suited for Faraday tomography. However, the quality of the Faraday depth cubes could be improved significantly by performing additional processing to remove the instrumental polarization leakage, for example through the use of complex CLEAN \citep{Pratley2016} to remove image-plane sidelobes. It may also be worthwhile to explore higher (angular) resolution, as this could increase sensitivity to emission with stronger Faraday depth gradients. At the time of this work, LOTSS data with direction-dependent calibration was not available for most observations. Early investigations into the effects of direction-dependent polarization on the polarization properties suggest that there is a modest improvement due to fewer image-plane sidelobes of the instrumental leakage around bright sources (S. O'Sullivan, private communication), which is encouraging for future polarization work with LOTSS data. LOTSS will observe the entire sky north of declination zero, so a Faraday tomography survey, with excellent sensitivity and resolution (both image-plane and in Faraday depth), will soon be possible.

\appendix
\section{Beam depolarization in a linear Faraday depth gradient}\label{sec:depolarization}
Gradients in Faraday depth with respect to position on the sky result in corresponding gradients in the polarization angle (if the intrinsic emitted angles are uniform across the source), which in turn produce beam depolarization when observed with a finite resolution. Here we explore the depolarization caused by Faraday depth gradients by considering the case of an infinite, uniform background emission source, with a (foreground) gradient in Faraday depth, corresponding to the sheets of emission and Faraday depth gradients we observe in our data. When a Gaussian beam is assumed, an analytical solution can be found, as described in \citet{Sokoloff98}. However, more realistic synthesized beams have not been considered in the literature, and below we show that this can make a significant difference in the expected depolarization.

Consider a gradient in Faraday depth with respect to an angular position variable $x$, $\frac{\mathrm{d}\phi}{\mathrm{d}x}$. The resulting intrinsic (sky) polarization, in the complex-polarization notation\footnote{We use a tilde to denote complex quantities, specifically phasors describing linear polarization.}, is 
\begin{equation}\label{eq:sky}
\tilde{P}_\mathrm{sky}(x,y,\lambda) = \tilde{P}_0 \, \exp\left(2i \lambda^2 \frac{\mathrm{d}\phi}{\mathrm{d}x} x\right)
\end{equation}
where $\lambda$ is the observing wavelength and $\tilde{P}_0$ is the intrinsic emitted polarization prior to Faraday rotation.\footnote{For convenience we have aligned the gradient with the x-axis and set $\phi$=0 at $x$=0, but these are arbitrary and do not affect the final result.} The observed polarization is the convolution of the intrinsic polarization with the telescope beam. In the case of a circular Gaussian beam ($\frac{1}{2\pi\sigma^2}\exp({-\frac{x^2+y^2}{2\sigma^2}})$), the convolution integral can be solved analytically, resulting in the previously known result
\begin{equation} \label{eq:gauss}
\tilde{P}_\mathrm{obs}(x,y,\lambda) = \tilde{P}_0 \, \exp\left(2i \lambda^2 \frac{\mathrm{d}\phi}{\mathrm{d}x} x\right) \; \exp\left(-2\left( \lambda^2 \frac{\mathrm{d}\phi}{\mathrm{d}x} \sigma \right)^2\right).
\end{equation}

The first part of this equation is identical to the intrinsic polarization and contains all the position dependence and the only complex terms, which means that the correct polarization angle, Faraday depth, and Faraday depth gradient are recovered, but the polarized intensity is modified by the later parts of the equation. The second part of the equation gives the depolarization, and shows that the depolarization is a very strong function of the wavelength, gradient, and beam size.

However, a Gaussian is often not an accurate representation of the synthesized beam, depending the ($u,v$) coverage of the instrument used. Previous calculations such as \citet{Tribble91}, while not for the exact same model, have shown that Gaussian functions tend to cause much stronger depolarization than other functions. With this motivation, we have developed a more general formulation for beam depolarization of gradients.

The general case, and the more natural formulation for interferometric observations, can be found simply by exploiting properties of Fourier transforms. Since the observed polarization is the convolution of the sky polarization with the telescope beam, we can move to the Fourier domain where the convolution becomes a multiplication and the beam becomes the ($u,v$) coverage of the observation:
\begin{eqnarray}
\mathcal{F} \{ \tilde{P}_\mathrm{obs}(x,y) \} &=& \mathcal{F} \{ \tilde{P}_\mathrm{sky}(x,y) \ast \mathrm{Beam}(x,y) \}\\ 
&=& \mathcal{F} \{ \tilde{P}_\mathrm{sky}(x,y) \} \, \mathcal{F} \{ \mathrm{Beam}(x,y) \}\\
&=& \tilde{P}_0 \,\delta\left(u - \frac{2\lambda^2}{2\pi}\frac{\mathrm{d}\phi}{\mathrm{d}x}\right)\delta(v) \, W(u,v)
\end{eqnarray}
where $W(u,v)$ is the sampling weight function in the $(u,v)$ plane, and in the last step we have used the definition of $\tilde{P}_\mathrm{sky}(x)$ from Eq.~\ref{eq:sky} and recalled that the Fourier transform of a complex exponential is a suitably shifted Dirac delta function. 
Inverting the Fourier transform to recover the observed sky distribution gives:
\begin{eqnarray}
\tilde{P}_\mathrm{obs}(x,y) &=& \mathcal{F}^{-1} \left\{ \tilde{P}_0 \, \delta\left(u - \frac{2\lambda^2}{2\pi}\frac{\mathrm{d}\phi}{\mathrm{d}x}\right)\delta(v) \, W(u,v) \right\} \\
&=& \tilde{P}_0 \exp\left(2i \lambda^2 \frac{\mathrm{d}\phi}{\mathrm{d}x} x\right) \, W\left(\frac{\lambda^2}{\pi}\frac{\mathrm{d}\phi}{\mathrm{d}x},0\right).\label{eq:result}
\end{eqnarray}

To this point, the result is general and works for any sampling function; using Gaussian weights gives the result from Eq. \ref{eq:gauss}. 
The correct Faraday depth, and by extension the gradient in Faraday depth, is recovered at all positions, but the polarized intensity is modified by the weighting function at the location of the gradient in the ($u,v$) plane. As a result, gradients at positions in the ($u,v$) plane that are not sampled by the instrument or that have been down-weighted significantly will have zero or strongly reduced polarized intensity and may not be detected; for such an idealized gradient the beam depolarization depends only on the weight function.

The steepest gradient that can be measured for a given observation can be determined from three parameters: the observing wavelength, $\lambda$, the largest sampled $(u,v)$ coordinate, UV$_{\mathrm{max}}$ (normalized by wavelength), and the Faraday depth gradient, $\frac{\mathrm{d}\phi}{\mathrm{d}x}$. We found that we could reduce this to one parameter by converting our model into scale-free units: we replaced the UV$_{\mathrm{max}}$ parameter with the synthesized beam full-width half max (FWHM), which is defined as FWHM$=\mathrm{UV_{max}}^{-1}$; replaced the $\lambda^2 \frac{\mathrm{d}\phi}{\mathrm{d}x}$ term in the intrinsic polarization with a single variable, the polarization angle gradient  (with units of radians of rotation per beam FWHM); and expressed the position variable $x$ in terms of the FWHM. The resulting condition for being observable is $\frac{\mathrm{d}(\phi\lambda^2)}{\mathrm{d}x} \, \mathrm{FWHM} < \pi $, or equivalently less than a 180\degr\ change in polarization angle across the FWHM of the synthesized beam.

To derive equation \ref{eq:result}, we assumed that the source of the polarized emission has an infinite extent, so that it shows up in the ($u,v$) plane at a single point. Real sources, which have a finite extent, cover an area in the {\it uv}-plane, not a single point. A more general argument can be constructed from Fourier properties, using the ideal case as a guide. Consider a polarized intensity feature of arbitrary morphology and angular size (and its corresponding Fourier transform to the $(u,v)$ plane), and apply a linear Faraday depth gradient of the form in Eq.~\ref{eq:sky} which has the effect of adding a position dependent phase modulation. From the Fourier shift-modulation theorem, the net effect is to shift the Fourier transform of the polarized feature in the $(u,v)$ plane, with the length and orientation of the shift determined by the steepness and orientation of the Faraday depth gradient. This shift can cause some or all of the polarized feature to fall outside of the $(u,v)$ coverage of the observation, causing depolarization.

This result mirrors that of \citet{Schnitzeler2009}, who used the same gradient model and Fourier properties to derive the corresponding result for the shallowest gradient that can be detected, given the shortest baseline present in an observation. These can be interpreted together: the longest baseline sets the steepest Faraday depth gradient that can be observed, while the shortest baseline sets the shallowest gradient that can be observed. Single-dish observations, which measure around the center of the $(u,v)$ plane, are sensitive to large polarized features with small Faraday depth gradients.

It should also be noted that this derivation describes beam depolarization in a single frequency channel. This depolarization may have a very strong frequency dependence; as frequency decreases the location of the gradient in the $(u,v)$ plane will move outwards to larger $(u,v)$ distances, scaling as the square of the wavelength, while a baseline of a fixed length moves inwards proportionally to the wavelength. For a given observation there will be a range of gradients that can be detected in a fraction of the observed bandwidth, producing a complex frequency-dependence which may introduce artificial structure into the corresponding Faraday depth spectrum. Determining the specific effects of a `partial detection' of a gradient is beyond the scope of this derivation.

Overall, we find three important results from this derivation. First, we have identified a key figure of merit for characterizing Faraday depth gradients for the purpose of evaluating their depolarization within a given observation: the change in polarization angle across the FWHM of the synthesized beam, with gradients steeper than 180\degr\ per beam FWHM being beyond the $(u,v)$ coverage of the observation. 

Second, beam depolarization may often not be as strong as may have been previously predicted using a Gaussian beam model. Just below the threshold of 180\degr\ per FWHM the depolarization of an ideal gradient by the Gaussian beam would leave only 2.9\% of the original polarized flux, whereas a uniform-weighted observation with the same beam FWHM would not be depolarized. The effects will be less extreme for a more realistic gradient, depending on the weighting of the $(u,v)$ coverage of the observation and the angular extent of the gradient.

Third, and perhaps most importantly, this suggests that the choice of $(u,v)$ weighting in the imaging process can have very strong effects on the measured polarization. A full study of this effect is beyond the scope of this work, but these simple models imply that the use of Gaussian tapers and similar weighting schemes that highly down-weight longer baselines may cause significant depolarization; the use of such tapers should be considered cautiously to avoid removing possible polarized features of interest.

\begin{acknowledgements}
CVE would like to thank Justin Bray and Jeroen Stil for their helpful comments on the derivation in the appendix.\\

This work is part of the research program 639.042.915, which is (partly) financed by the Netherlands Organisation for Scientific Research (NWO).\\

LOFAR, the Low Frequency Array designed and constructed by ASTRON, has facilities in several countries, that are owned by various parties (each with their own funding sources), and that are collectively operated by the International LOFAR Telescope (ILT) foundation under a joint scientific policy.\\
This research used ionospheric TEC maps produced by the Centre for Orbit Determination in Europe (CODE, http://aiuws.unibe.ch/ionosphere/). Some of the ISM tracer maps were downloaded from the Legacy Archive for Microwave Background Data Analysis (LAMBDA, https://lambda.gsfc.nasa.gov/), part of the High Energy Astrophysics Science Archive Center (HEASARC). HEASARC/LAMBDA is a service of the Astrophysics Science Division at the NASA Goddard Space Flight Center..

This research made extensive use of Astropy, a community-developed core Python package for Astronomy \citep{Astropy}; SciPy \citep{Scipy}; NumPy \citep{Numpy}; IPython \citep{Ipython}; matplotlib \citep{Matplotlib}; the Common Astronomy Software Applications \citep[CASA,][]{CASA}; and the Karma visualization tools \citep{Karma}.
\end{acknowledgements}

\bibliographystyle{aa} 
\bibliography{References} 

\end{document}